\documentclass[article]{JHEP3}

\usepackage{epsfig,multicol,amsmath}
\newcommand\fverb{\setbox\pippobox=\hbox\bgroup\verb}
\newcommand\fverbdo{\egroup\medskip\noindent%
\fbox{\unhbox\pippobox}\ }			
\newcommand\fverbit{\egroup\item[\fbox{\unhbox\pippobox}]}
\newbox\pippobox
\def\d2bar{$\overline{\mbox D2}$}
\title{String Pair Creations in D-brane Systems}
\author{Jin-Ho Cho, Phillial Oh, Cheonsoo Park, and Jonghyeon Shin\\
BK21 Physics Research Division and Institute of Basic Science\\
 Sungkyunkwan University, Suwon 440-746, Korea\\
E-mail: \email{jhcho@newton.skku.ac.kr}, \email{ploh@dirac.skku.ac.kr}, \email{cspark@newton.skku.ac.kr}, \email{gempyonn@korea.com}}

\preprint{\hepth{0501190}} 

\abstract{We investigate the criterion, on the Born-Infeld background fields, for the open string pair creation to occur in D$p$-(anti-)D$p$-brane systems. Although the pair creation occurs generically in both D$p$-D$p$ and D$p$-anti-D$p$ systems for the cases which meet the criterion, it is more drastic in D$p$-anti-D$p$-brane systems by some exponential factor depending on the background fields. Various configurations exhibiting pair creations are obtained via duality transformations. These include the spacelike scissors and two D-strings (slanted at different angles) passing through each other. We raise the scissors paradox and suggest a resolution based on the triple junction in IIB setup.}

\keywords{pair creation, D-brane, open string, D-$\overline{\text{D}}$ system}

\begin{document} 
\section{Introduction and Summary}

\subsection{Historical Overview}

Pair creation of particles in the background of strong electric field \cite{schwinger51} is one of the most fascinating features discerning quantum field theory from quantum mechanics. The pair creation rate (of particles with spin $J=0$ or $1/2$, charge $\pm q$, and mass $M$), is given by the imaginary part of the vacuum energy density; 
\begin{eqnarray}\label{schwinger}
\omega & = & \frac{2J+1}{8\pi^3}\sum^{\infty}_{l=1}\left(\frac{qE}
{l}\right)^{2}(-1)^{(l+1)(2J+1)}e^{-\frac{\pi l M^2}{|qE|}},
\end{eqnarray}
where $E$ is the external electric field. The virtual pair due to the vacuum fluctuation becomes separated to be on-shell by the effective potential barrier formed by the external electric field \cite{Brout:1995}. 

It is natural to expect analogous physics in the framework of the string theory because the electric field is attainable in the low energy limit $(\alpha'\rightarrow 0)$ from the open string modes and the open string carries corresponding charges at its end points. The first attempt to realize the string pair creation was made by Bachas and Porrati \cite{bachas92}. They considered the unoriented string in a background of constant electric field. Among the one loop diagrams (the annulus, M\"obius strip, and the torus),  the annulus diagram turned out to be relevant and the result approaches Schwinger's field theory result (\ref{schwinger}), in the weak electric field limit (in the string unit). Pair creation of open strings happens generically {\it unless they are neutral} (having zero net charge).  

\subsection{Questions}

With the new ingredient of D-branes \cite{Polchinski:1995} in the string theory, one can ask about the pair creation of open strings over various D-branes in type-IIA or type-IIB setup. One virtue of thinking of open string pair creation in D-brane systems is that various dual transformations provide us with new insights on many phenomena which are hard to be recognized otherwise. Naively, the oriented open strings coupled to IIA- or IIB-closed strings, carrying zero net charge, do not seem to be pair-created. However we show, in this paper, that the open strings stretched between a $\text{D}p$- and another parallel (anti-)$\text{D}p$-brane (hereafter we will use anti-$\text{D}p$ and $\overline{\text{D}p}$ interchangeably) can be created in pair when different electric fields are appropriately arranged on each of the branes.   

Another motivation of this paper is concerned with the tachyon modes present generically over D-$\overline{\text{D}}$-brane system for a specific combination of Born-Infeld electric and magnetic fields. In the authors' previous paper \cite{cho03}, the necessary condition, for BPS configurations of a D$2$-$\overline{\text{D}2}$ system with the electric- and magnetic-fields, was obtained;
\begin{eqnarray}\label{notachyon}
D\equiv\left(1-\vec{e}\cdot\vec{e}+\vec{b}\cdot\vec{b}\right)\left(1-\vec{e'}\cdot\vec{e'}+\vec{b'}\cdot\vec{b'}\right)-\left(1-\vec{e}\cdot\vec{e'}+\vec{b}\cdot\vec{b'}\right)^2=0,
\end{eqnarray}
where $(\vec{e},\,\vec{b})$ and $(\vec{e'},\,\vec{b'})$ are sets of the electric and the magnetic fields over each (anti-)D2-brane. When the value $D$ in the above equation is positive, the tachyon mode survives GSO projection. In this paper, we study the case where the value is negative. We will see that this case is characterized by the imaginary modes, which cause the pair creation. Especially we emphasize that the analysis assumes the most general arrangement of non-parallel electric fields, which provides us with various dual physics including `Hanany-Witten'-like effect \cite{Hanany:1997} for two D-strings passing through each other.

Concerning previous motivation, one can also ask about the role of the magnetic fields in pair creations. This is especially important in relation with the question of the electron-positron pair creation around the pulsar magnetosphere \cite{Sturrock:1971,Tsai:1974}. In the framework of quantum field theory, this question was already answered in the original paper \cite{schwinger51}, where the pure magnetic case was shown to give only real value of the vacuum energy density, thus no pair creation occurs \footnote{It is possible to make pairs by scattering photons off a strong magnetic field source \cite{Tsai:1974}.}. (See also a recent argument in Ref. \cite{kim00}.) However one can see, in the T-dual description, at least the subsidiary role of the magnetic fields. For example, different values of magnetic fields over D$2$- and $\overline{\text{D}2}$-branes result in different tilting angles of D-strings in the T-dual setup. The negativity of the value $D$ in Eq. (\ref{notachyon}) corresponds to the super-luminal speed of the intersection of two D-strings. These spacelike scissors are unstable as shown in \cite{bachas02}, and our configurations include their T-dual counterpart.

\subsection{Scheme and Results}

Let us briefly summarize the scheme and the results of this paper. The earlier part is about the quantization of the open string stretched between a D$2$-brane and another parallel (anti-)D$2$-brane (`inter-string' in short), for a general background of the electric and the magnetic fields causing the imaginary modes of the string. In the next section, we explain the basic setup and notations used in this paper. By comparing the boundary conditions, we observe the correspondence between our configuration and that of Ref. \cite{bachas92}. In the D-brane language, the configuration studied in Ref. \cite{bachas92} can be interpreted as a pair of D-branes with parallel or anti-parallel electric fields on each brane. We consider more general situation allowing for non-parallel electric fields and different magnetic fields on those branes. In this general setup, we obtain the criterion, on the background fields, for the open string pair creation to occur. We emphasize the fact that for an appropriate arrangement of the background Born-Infeld gauge fields, the boundary conditions at two ends of the inter-string shift the mode frequencies (of one temporal and one spatial components) by some imaginary part, which is determined by the background fields. In Sec. \ref{seciii}, we write down some relevant commutators among the integer or imaginary modes and defer the details about quantization to App. \ref{appi}. Subsequently, we showed that the kinetic momentum of the inter-string is space-like as a consequence of the removal of one temporal and one spatial components with integer mode. 

The later part of the paper discusses the pair creation. In Sec. \ref{seciv}, we compute the annulus diagram for the inter-string in an arbitrary D$p$-D$p$ or D$p$-$\overline{\text{D}p}$ brane system. Sec. \ref{secv} rewrites the one-loop amplitude in a simpler form in terms of mass spectrum. In both D$p$-D$p$ and D$p$-$\overline{\text{D}p}$ brane systems, pair creations occur when the criterion obtained in the earlier part is satisfied. However the pair creation is more drastic in D$p$-$\overline{\text{D}p}$ brane systems. For example in the weak field limit, the pair creation is enhanced by the factor $\exp{(\pi^2/|\vec{e}-\vec{e'}|)}$ in D$p$-$\overline{\text{D}p}$ brane systems, compared to the D$p$-D$p$ systems. Sec. \ref{secvi} shows various dual configurations exhibiting pair creations. We first show how our result corresponds to that of Schwinger \cite{schwinger51} and Bachas and Porrati \cite{bachas92} in a proper case of background fields. After a series of duality transformation, the case with the orthogonal electric fields ($\vec{e}\bot\vec{e'}$) shows a behavior similar to (but not the same as) Hanany-Witten effect \cite{Hanany:1997}; two D-strings (slanted at different angles) generate a pair of open strings between them when they `pass through' each other. The case with the anti-parallel electric fields and non-vanishing magnetic fields is dual to the space-like scissors. This latter case actually raises the question of scissors paradox because the inter-strings keep up with the super-luminal intersection. The main reason causing this puzzling results is the rigid boundary ansatz for D-branes. We propose a resolution based on the triple junction formed at the point where the inter-string meets D-string. The tension balance at the junction forbids the inter-string to flee at the super-luminal speed.

\section{Setup}

\subsection{Configuration and Notations}

We will focus on D$2$-(anti-)D$2$ case. Higher dimensional extension can be obtained via appropriate T-dualities normal to D$2$ directions. 
In the notion of the fact that the open string stretched between two parallel D$2$- and $\overline{\text{D}2}$-branes can make only a lineal motion along the direction normal to $\vec{e}-\vec{e'}$ \cite{cho03}, it is convenient to configure the whole system as in Fig. 1, so that the string move along the $X^{2}$-axis. Each end of the inter-string is coupled to different background Born-Infeld fields, $B^{(\sigma)}_{\mu\nu}$ $(\sigma=0,\,\,\pi)$ on each brane world-volume, which are 
\FIGURE{
\epsfbox{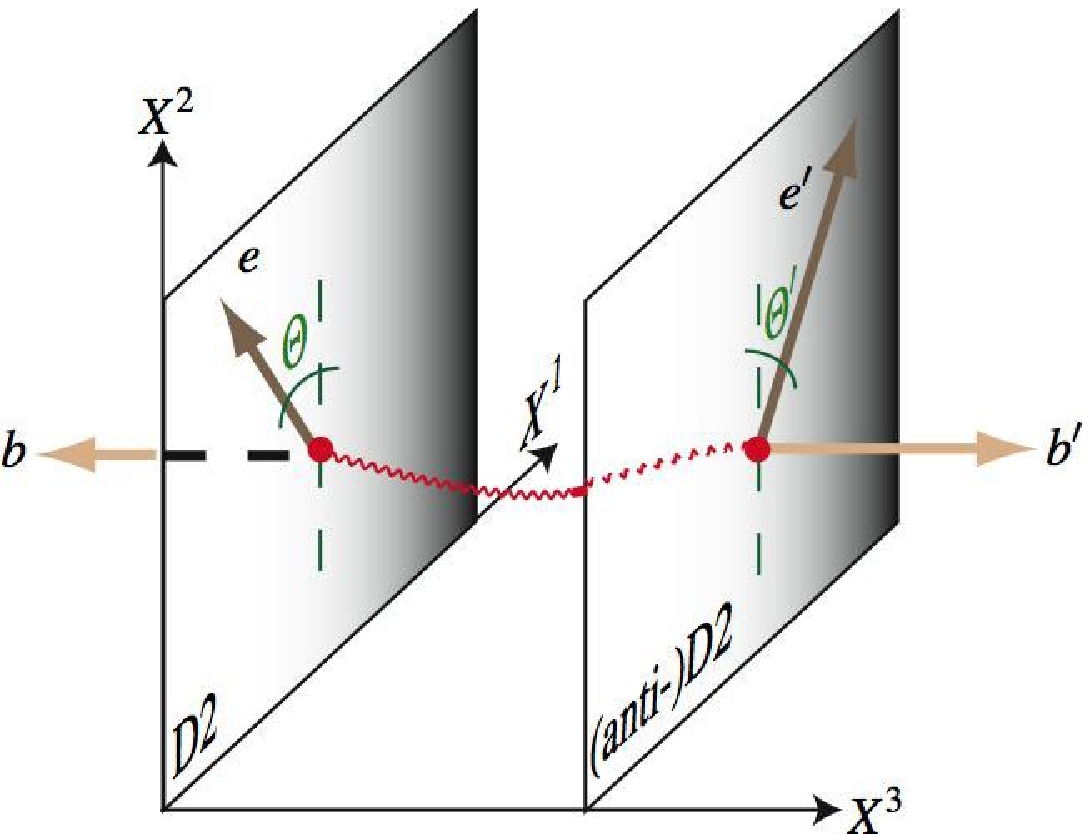}
\caption{An open string is stretched between a D$2$-(anti-)D$2$-pair. Each end point is coupled to the Born-Infeld electric- and magnetic-field on each brane. The string can move along $X^{2}$-axis.}
}
written by 
\begin{equation}
\left(B^{(0)}_{\mu\nu}\right) =
\left( \begin{array}{ccc}
0  & e\sin\theta &  -e\cos\theta \\
-e\sin\theta  & 0 & -b \\
e\cos\theta & b & 0
\end{array} \right) \,,
~~~~~~~~~~
\left(B^{(\pi)}_{\mu\nu}\right) =
\left( \begin{array}{ccc}
0 & -e^{\prime}\sin\theta' & -e^{\prime}\cos\theta' \\
e^{\prime}\sin\theta' & 0 & b^{\prime} \\
e^{\prime}\cos\theta' & -b^{\prime} & 0
\end{array} \right) \,,
\end{equation}
where $\theta+\theta'$ is the angle between $\vec{e}$ and 
$\vec{e}'$ and $e\cos\theta = e^{\prime}\cos\theta'$ (so that $\vec{e}-\vec{e'}$ is normal to $X^{2}$). In the following, the expression $E^{(\sigma)}_{\mu\nu}\equiv \eta_{\mu\nu}+B^{(\sigma)}_{\mu\nu}$ will also be frequently used. Note that all the primed letters pertain to the brane at $\sigma=\pi$. Actually a proper Lorentz transformation could be taken to remove both magnetic fields $b$ and $b'$ at the same time. However, we leave them alive as regulating parameters to deal with some delicate situations, which will be discussed in later sections.

\subsection{Equations of Motion and Boundary Conditions}

In the flat space-time, the Neveu-Schwarz (NS) superstring is described by the action;
\begin{equation}
S = -\frac{1}{4\pi\alpha^{\prime}}  \int d^2\sigma \left [
\begin{array}{ll} 
\partial_{\alpha} X^{\mu}
\partial^{\alpha} X^{\nu} \eta_{\mu\nu} +i \bar{\psi}^{\mu}\gamma^{\alpha}
\partial_{\alpha}\psi^{\nu} \eta_{\mu\nu}\\
{}&{}\\\displaystyle
{}+ \Bigl( 2A_{\mu}^{(\sigma)} \dot{X}^{\mu} 
+ \frac{i}{2} \bar{\psi}^{\mu}\gamma^{1}\psi^{\nu} B^{(\sigma)}_{\mu\nu} \bigr) 
\bigl( \hat{\delta}(\sigma)-\hat{\delta}(\sigma-\pi) \Bigr) 
\end{array}
\right] \,,
\end{equation}
where gamma matrices are related to Pauli matrices as $\gamma^{0}=i\sigma^{2}$, $\gamma^{1}=\sigma^{1}$. The relative minus sign in the boundary terms at $\sigma=0$ and $\sigma=\pi$ implies that the two ends are oppositely charged.
We give mixed type boundary conditions along the brane world-volume directions and Dirichlet type for the residual directions. Especially we assume the distance, $y$, between D$2$- and (anti-)D$2$-branes. In the light-cone coordinates $\sigma^{\pm}=\tau\pm\sigma$, they are 
\begin{eqnarray}\label{bcboson}
&& \Bigl( E^{(\sigma)}_{\nu\mu}\partial_{+} X^{\nu} - E^{(\sigma)}_{\mu\nu}
\partial_{-} X^{\nu} \Bigr) \Big{|}_{\sigma=0,\pi}  = 0 \,, 
~~~~~~~~~~ (\mu , \nu=0, 1, 2) \,, \nonumber\\
&& X^3 |_{\sigma =0} = 0 \,,~~~~~ X^3 |_{\sigma =\pi} =y \,, \nonumber\\
&& X^{\mu} |_{\sigma= 0,\pi} = 0 \,,
~~~~~~~~~~ \qquad \qquad(\mu = 4 , \cdots , 9)\,.
\end{eqnarray}
As for the fermions, there are two possible consistent boundary conditions corresponding to Ramond (R)- and Neveu-Schwartz (NS)-sector respectively, 
\begin{eqnarray}
&&\left\{
\begin{array}{l}
E^{(\sigma)}_{\nu\mu}\psi^{\nu}_{+} |_{\sigma =0}
  = E^{(\sigma)}_{\mu\nu}\psi^{\nu}_{-} |_{\sigma =0} \,, \\
E^{(\sigma)}_{\nu\mu}\psi^{\nu}_{+} |_{\sigma =\pi}
  =\pm E^{(\sigma)}_{\mu\nu}\psi^{\nu}_{-} |_{\sigma =\pi} \,,
\end{array}\right.
~~~~~~~~~~ (\mu , \nu = 0, 1, 2) \,, \nonumber\\
&& \left\{
\begin{array}{l}
\psi^{\mu}_{+} |_{\sigma =0} =-\psi^{\mu}_{-} |_{\sigma =0} \,, \\
\psi^{\mu}_{+} |_{\sigma =\pi} =\mp\psi^{\mu}_{-} |_{\sigma =\pi} \,,
\end{array}\right.
~~~~~~~~~~\qquad\qquad (\mu = 3,4, ... ,9) \,,
\end{eqnarray}
where the upper/lower sign is for the R-/NS-sector and $\psi^{\mu}=(\psi^{\mu}_{+}~,~\psi^{\mu}_{-})$ is the Majorana spinor.

Comparing the boundary condition (\ref{bcboson}) with that of Ref. \cite{bachas92}, where the unoriented open string with arbitrary charges $q_{1},\,q_{2}$ at its two ends was considered to be in a constant electric background $F_{i0}$, we note the following conversion relations;
\begin{eqnarray}\label{bachas}
B^{(0)}_{i0} & = & \pi q_{1}F_{i0}\nonumber \\
B^{(\pi)}_{i0} & = & -\pi q_{2}F_{i0}. 
\end{eqnarray}
In the case of opposite charges in \cite{bachas92}, that is when $q_{1}+q_{2}=0$, there is no pair production. In our language, this corresponds to the case of two parallel D-branes with the same electric fields, $B^{(0)}_{i0}= B^{(\pi)}_{i0}$, on both world volumes, which is obviously a BPS configuration \cite{cho03}.
  
\subsection{Relations among the Mode Coefficients}

For the time being, we pay attention to the first three nontrivial directions of $X^{\mu}$, $(\mu=0,\,1,\,2)$. The boundary condition at $\sigma=0$ regulates them to the following expression, 
\begin{eqnarray}\label{solbo}
X_{\mu} &=&
\frac{x_{\mu}}{2} + E^{(0)}_{\mu\nu} \left[  a_{0}^{\nu}\sigma^{+} 
+ \displaystyle\sum_{n\ne0}^{\infty}\frac{i}{n}a_{n}^{\nu}e^{-in\sigma^{+}}+
\right.\nonumber\\&&\qquad 
\left. + \displaystyle\sum_{n=-\infty}^{\infty}\frac{i}{2(n+\kappa)} 
\Big( a_{n+\kappa}^{\nu}e^{-i(n+\kappa)\sigma^{+}} 
- \bar{a}_{-n-\kappa}^{\nu} e^{i(n+\kappa)\sigma^{+}} \Big) \right]\nonumber\\
&+& \frac{x_{\mu}}{2} + E^{(0)}_{\nu\mu} \left[ a_{0}^{\nu}\sigma^{-} 
+ \displaystyle\sum_{n\ne0}^{\infty}\frac{i}{n}a_{n}^{\nu}e^{-in\sigma^{-}}+
\right.\nonumber \\&&\qquad 
\left.+ \displaystyle\sum_{n=-\infty}^{\infty}\frac{i}{2(n+\kappa)} 
\Big( a_{n+\kappa}^{\nu}e^{-i(n+\kappa)\sigma^{-}} 
- \bar{a}_{-n-\kappa}^{\nu} e^{i(n+\kappa)\sigma^{-}} \Big) \right]
 \,.
\end{eqnarray}
The non-integer mode part, as we allowed in the general form, is non-trivial if the boundary conditions at $\sigma=\pi$ are different from those at $\sigma=0$. 

In order for the above expressions to satisfy the boundary conditions at $\sigma=\pi$, the coefficients are constrained so that
\begin{eqnarray}
&&a_{n}^{1} = 0, \qquad (e\sin\theta+e^{\prime}\sin\theta')\, a_{n}^{0} = -(b+b')\,a_{n}^{2} \,,
\label{in}
\end{eqnarray}
and 
\begin{eqnarray}
&&\bigg[i\tan(\kappa\pi)\Big(\eta_{\mu\nu} - B^{(\pi)}_{\mu\rho}
\eta^{\rho\lambda} B^{(0)}_{\lambda\nu}\Big) - \Big(B^{(0)}_{\mu\nu} 
- B^{(\pi)}_{\mu\nu}\Big) \bigg] ~ a_{n+\kappa}^{\nu} = 0 \,, \nonumber\\
&&\bigg[i\tan(\kappa\pi)\Big(\eta_{\mu\nu} - B^{(\pi)}_{\mu\rho}
\eta^{\rho\lambda} B^{(0)}_{\lambda\nu}\Big) + \Big(B^{(0)}_{\mu\nu} 
- B^{(\pi)}_{\mu\nu}\Big) \bigg] ~ \bar{a}_{-n-\kappa}^{\nu} = 0 \,.
\label{ep}
\end{eqnarray}
Eq. (\ref{ep}) has nontrivial solutions provided that the determinant of the pre-factor matrix vanishes, that is when
\begin{equation}\label{beta}
\beta^{2}\equiv \Big(\tan(\kappa\pi)\Big)^{2} 
= \frac{\Big(1-\vec{e}\cdot\vec{e}+\vec{b}\cdot\vec{b}\Big)\Big(1-\vec{e'}
\cdot\vec{e'}+\vec{b'}\cdot\vec{b'}\Big) 
- \Big(1-\vec{e}\cdot\vec{e'}+\vec{b}\cdot\vec{b'}\Big)^{2}}
{\Big(1-\vec{e}\cdot\vec{e}^{\prime}+\vec{b}\cdot\vec{b'}\Big)^{2}} \,.
\end{equation}
Here we used the vector notation, $\vec{e}=\Big(B^{(0)}_{10}~,~B^{(0)}_{20}~,~0\Big)$, 
$\vec{b}=\Big(0~,~0~,~B^{(0)}_{12}\Big)$, 
$\vec{e'}=\Big(B^{(\pi)}_{10}~,~B^{(\pi)}_{20}~,~0\Big)$, and 
$\vec{b'}=\Big(0~,~0~,~B^{(\pi)}_{12}\Big)$.

The same equation was obtained in Ref. \cite{cho03} despite of different coordinate systems. The authors assumed in the paper that $\beta^{2}\geq 0$ and got the result that the tachyon mode disappear only when $\kappa=0$. In this paper, we consider the cases where $\beta^{2}< 0$. It is convenient to introduce a real parameter $\epsilon=-i\kappa$\,\footnote{Without loss of generality, one can assume $\epsilon>0$.}. Then
\begin{equation}
\label{epsilon}
\tanh^{2}(\epsilon\pi)=-\beta^{2}
\end{equation}
The reality condition on the fields $X^{\mu}$ $(\mu=0,\,1,\,2)$ relates the coefficients with their complex conjugates as
$a_{-n}^{\nu}=(a_{n}^{\nu})^{\ast}$, 
$a_{-n+i\epsilon}^{\nu}=(a_{n+i\epsilon}^{\nu})^{\ast}$, 
and $\bar{a}_{n-i\epsilon}^{\nu}=(\bar{a}_{-n-i\epsilon}^{\nu})^{\ast}$. 

Due to the Eq. (\ref{ep}), $a_{n+i\epsilon}^{\nu}$ and $\bar{a}_{-n-i\epsilon}^{\nu}$ are not independent either. 
\begin{eqnarray}\label{in2}
&&a^{1}_{n+i\epsilon}=\frac{r+is}{p+iq} ~ a^{0}_{n+i\epsilon} \,, \quad ~~~~~~~~~~ 
a^{2}_{n+i\epsilon}=\frac{t+iu}{p+iq} ~ a^{0}_{n+i\epsilon} \,, \nonumber\\
&&\bar{a}^{1}_{-n-i\epsilon}=\frac{r-is}{p-iq} ~ \bar{a}^{0}_{-n-i\epsilon} \,, ~~~~~~~~~
\bar{a}^{2}_{-n-i\epsilon}=\frac{t-iu}{p-iq} ~ \bar{a}^{0}_{-n-i\epsilon} \,,
\label{fr}
\end{eqnarray}
where
\begin{eqnarray}
p &=& be^{\prime}\beta^{2}\cos\theta'(1-bb^{\prime} 
-ee^{\prime}\cos(\theta+\theta')) \,, \nonumber\\
q &=& \beta\big[ e^{\prime}(1+b^{2}-ee^{\prime}\cos\theta\cos\theta')
\sin\theta' + e(1-bb^{\prime}-ee^{\prime}\cos\theta\cos\theta')
\sin\theta\big] \,, \nonumber\\
r &=& \beta^{2}(1-ee^{\prime}\cos\theta\cos\theta')(1-bb^{\prime}
-ee^{\prime}\cos(\theta+\theta')) \,, \nonumber\\
s &=& 0 \,, \nonumber\\
t &=& -ee^{\prime}\beta^{2}\sin\theta\cos\theta'(1-bb^{\prime}
-ee^{\prime}\cos(\theta+\theta')) \,, \nonumber\\
u &=& \beta\big[(b+b^{\prime})(-1+ee^{\prime}\cos\theta\cos\theta')
- bee^{\prime}\sin\theta\sin\theta' + e^{2}b^{\prime}\sin^{2}\theta
\big] \,.
\end{eqnarray}

Let us move on to the fermionic part. The solutions $\psi_{\pm}^{\mu}$, 
satisfying the boundary condition at $\sigma=0$, look very similar to the 
expressions of $\partial_{\pm}X^{\mu}$. The only difference is that the mode 
frequency $n$ is replaced by $r$, that is integer valued in the R-sector and 
half integer valued in the NS-sector;
\begin{eqnarray}\label{solfer}
\psi_{\mu+} = \sum_{r}E^{(0)}_{\mu\nu}\bigg[ h_{r}^{\nu}e^{-ir\sigma^{+}} 
+ \frac{1}{2}\Big( h_{r+i\epsilon}^{\nu}e^{-i(r+i\epsilon)\sigma^{+}} 
+ \bar{h}_{-r-i\epsilon}^{\nu}e^{i(r+i\epsilon)\sigma^{+}} \Big) \bigg] \,,
\nonumber\\
\psi_{\mu-} = \sum_{r}E^{(0)}_{\nu\mu}\bigg[ h_{r}^{\nu}e^{-ir\sigma^{-}} 
+ \frac{1}{2}\Big( h_{r+i\epsilon}^{\nu}e^{-i(r+i\epsilon)\sigma^{-}} 
+ \bar{h}_{-r-i\epsilon}^{\nu}e^{i(r+i\epsilon)\sigma^{-}} \Big) \bigg] \,.
\end{eqnarray}
As with the
bosonic case, the fermionic part undergoes nontrivial relations among the mode 
coefficients upon the imposition of the boundary conditions on the other end of 
the string $(\sigma=\pi)$. The precise form of those dependencies turns out to 
be the same as the form shown in eq. (\ref{in}) and in eq. (\ref{fr}), but with 
the replacements of $a$'s with $h$'s.

\section{Quantization}\label{seciii}

\subsection{Symplectic Form and Mode Expansion}

The constraints (\ref{in}) and (\ref{fr}) on the oscillation modes of $\mu=0,1,2$ directions make the quantization quite involved. One simple way to get out of this nontrivial situation is to make use of the symplectic form; 
\begin{equation}
\Omega = \int_{0}^{\pi} d{\sigma}\,\delta\Pi_{X_{\mu}}\wedge\delta X^\mu 
- \int_{0}^{\pi} d{\sigma}\,\delta\Pi_{\psi_{\mu}}\wedge\delta\psi^{\mu} \,,
\end{equation}
where we denoted the exterior derivative in the phase manifold as '$\delta$'. The explicit forms of 
the conjugate momenta are 
\begin{eqnarray}
\Pi_{X_{\mu}} &=& \frac{1}{2\pi\alpha^{\prime}}\bigg[\eta_{\mu\nu}\partial_{\tau}
X^{\nu} - \Big(A_{\mu}^{(0)}\hat{\delta}(\sigma-0) - A_{\mu}^{(\pi)}\hat{\delta}
(\sigma-\pi)\Big)\bigg] \,, \nonumber\\
\Pi_{\psi_{\mu}} &=& \frac{i}{4\pi\alpha^{\prime}} \bar{\psi}^{\nu}\gamma^{0}
\eta_{\mu\nu} \,.
\end{eqnarray}

We first perform the integration to get the symplectic two form in the mode expanded fashion. The resulting expressions for the bosonic part $\Omega_{B}$ and the fermionic part are given by
$\Omega_{F}$ are
\begin{eqnarray}
\Omega_{B} &=&
-\frac{1}{4\pi\alpha^{\prime}}\Big(B^{(0)}-B^{(\pi)}\Big)_{\mu\nu}
\delta x^{\mu}\wedge\delta x^{\nu} - \frac{1}{\alpha^{\prime}}
\Big(\eta-B^{(\pi)}\eta^{-1}B^{(0)}\Big)_{\mu\nu}\delta x^{\mu}\wedge\delta 
a_{0}^{\nu} \nonumber\\
&&+ \displaystyle\sum_{n\ne0}\frac{i}{n\alpha^{\prime}}\Big(\eta-B^{(0)}
\eta^{-1}B^{(0)}\Big)_{\mu\nu} \delta a_{-n}^{\mu}\wedge\delta a_{n}^{\nu} \nonumber\\
&&+ \displaystyle\sum_{n}\frac{i}{2(n+i\epsilon)\alpha^{\prime}}\Big(\eta-B^{(0)}
\eta^{-1}B^{(0)}\Big)_{\mu\nu} \delta \bar{a}_{-n-i\epsilon}^{\mu}\wedge
\delta a_{n+i\epsilon}^{\nu}\,, \nonumber\\
\Omega_{F} &=& \sum_{r}\frac{i}{2\alpha^{\prime}}\Big(\eta-B^{(0)}\eta^{-1}
B^{(0)}\Big)_{\mu\nu} \bigg[\delta h_{-r}^{\mu}\wedge\delta h_{r}^{\nu} 
+ \frac{1}{2}\delta \bar{h}_{-r-i\epsilon}^{\mu}\wedge\delta h_{r+i\epsilon}^{\nu} 
\bigg] \,.
\label{sym}
\end{eqnarray}

We next work out the constraints completely so that the symplectic two form, written in terms of independent phase variables $q^{M}=\{ x^{0},$ $x^{1},$ $x^{2},$ $a_{n}^{2},$ $a_{n+i\epsilon}^{0},$ $\bar{a}_{-n-i\epsilon}^{0},$ $h_{r}^{2},$ $h_{r+i\epsilon}^{0},$ $\bar{h}_{-r-i\epsilon}^{0}\}$, be of the form $\Omega=\Omega_{MN}\,\delta q^{M}\wedge\delta q^{N}/2$. Then Poisson bracket (now Dirac bracket because no constraint is remaining) of the functions, $f(q)$ and $g(q)$, on the phase space, can be 
written as $\{ f~,~g \}=\Omega^{MN}\partial_{M}f\partial_{N}g$, where 
$\Omega^{MN}\Omega_{NL}=\delta^{M}_{L}$. The conventional replacement of Dirac brackets with the (anti-)commutators will lead us to the operator algebra. Since the results are a bit complicated, we defer the details to App. \ref{appi}. Instead here, we just summarize the final simplified form obtained after suitable normalization of the operators;  
\begin{eqnarray}
\left[\alpha_{m}~,~\alpha_{n}\right] = m\delta_{m+n} \,, ~~~~~~~~~~ 
\left[\bar{\alpha}_{m-i\epsilon}~,~\alpha_{n+i\epsilon}\right] 
= -(m-i\epsilon)\delta_{m+n} \,, \nonumber\\
\{\varphi_{r}~,~\varphi_{s}\} = \delta_{r+s} \,, ~~~~~~~~~~ 
\{\bar{\varphi}_{r-i\epsilon}~,~\varphi_{s+i\epsilon}\}
= -\delta_{r+s} \,,
\end{eqnarray}
where the integer modes and the non-integer modes pertain to the direction $X^2$ of the transverse dimension and the longitudinal direction, respectively.

\subsection{Spacelike Motion of the Inter-string}

As was shown in Ref. \cite{cho03}, the motion of the D-$\overline{\text{D}}$ string is not omnidirectional. The string rather moves along the direction normal to the vector $\vec{e}-\vec{e'}$. Under the gauge $A_{\mu}=-B_{\mu\nu}X^\nu$, the integration of ${\Pi_X}_\mu$ gives the momentum in terms of the bosonic zero mode $a^\mu_0$; 
\begin{eqnarray}\label{fullmomentum}
P_\mu&=&\frac{1}{2\pi\alpha'}\int^\pi_0d\sigma\,\,\partial_\tau X_\mu  -\frac{1}{2\pi\alpha'}\left(A^{(0)}_\mu-A^{(\pi)}_\mu\right)\nonumber\\
&=&\frac{1}{2\pi\alpha'}\left[2\pi\eta_{\mu\nu}a^\nu_0-2\pi B^{(\pi)}_{\mu\rho}\eta^{\rho\lambda}B^{(0)}_{\lambda\nu}a^\nu_0+ \left(B^{(0)}_{\mu\nu}-B^{(\pi)}_{\mu\nu}\right)x^\nu\right].
\end{eqnarray}
The first term in the second line, coming from $\partial_\tau X_\mu$, corresponds to the kinetic momentum $k_\mu$, which differs from the conjugate momentum $P_\mu$ in the canonical way in the presence of gauge fields. The second and the third terms in the same line describe the momentum acquired from background fields, thus make the total momentum dependent on the space-time mean position $x^\mu$ of the string. The oscillatory parts of the kinetic momentum and the field momentum nicely cancel.
Thanks to the relation (\ref{in}) among the zero modes, one can write the kinetic momentum in terms of the component $a^2_0$ only;
\begin{eqnarray}\label{momentumvec}
\left(k^\mu\right)=\frac{a^2_0}{\alpha'}\left(-\frac{b+b'}{e\sin\theta+e'\sin{\theta'}},\,0,\,1\right).
\end{eqnarray}

It is interesting to see that this kinetic part is spacelike for the field configurations (with $\beta^2<0$) we are discussing in this paper \footnote{This does not mean the total momentum is spacelike. In fact, the background fields contribute the temporal component of the total momentum. We thank C. Bachas for pointing this.}. In fact,
\begin{eqnarray}\label{spacelike}
k^{\mu}\eta_{\mu\nu}k^{\nu} & = &\frac{|\vec{e}-\vec{e'}|^{2}-|\vec{b}-\vec{b'}|^{2}}{|\vec{e}-\vec{e'}|^{2}}\left(\frac{a^{2}_{0}}{\alpha'}\right)^{2} \nonumber\\
 & = & -\frac{|\vec{e}-\vec{e'}|^{2}-|\vec{b}-\vec{b'}|^{2}}{2\alpha'\beta^{2}(1-\vec{e}\cdot\vec{e'}+\vec{b}\cdot\vec{b'})^{2}}(\alpha_{0})^{2} .
\end{eqnarray}
In the second equality, we used the normalized zero mode component $\alpha_0$ (see Eq. (\ref{normalization}) in the appendix). In later sections, we will show that the spacelike kinetic momentum implies the spacelike motion and will discuss its physical consequences in more details.

In the sequel, it is convenient to use the following definition for $\xi$.
\begin{eqnarray}
\label{momentum}
(\alpha_{0})^{2}&=&-\frac{2\alpha'\beta^{2}(1-\vec{e}\cdot\vec{e'}+\vec{b}\cdot\vec{b'})^{2}}{(\vec{e}-\vec{e'})^{2}}(k^{2})^{2}\nonumber\\
&\equiv&\frac{(k^{2})^{2}}{\xi^{2}}
\end{eqnarray}

\section{Annulus Diagram}\label{seciv}

In quantum field theory, pair creation is characterized by the imaginary part of one-loop vacuum energy density;
\begin{eqnarray}
\cal{F}&=&\frac{i}{V_{D}}\ln Z \nonumber\\
& = & -\frac{i}{2V_{D}}\text{Tr} \ln(k_{\mu}k^{\mu}+m^{2})\nonumber \\
 & = & \frac{1}{2}\int \frac{d\vec{k}}{(2\pi)^{D-1}}\sqrt{\vec{k}\cdot\vec{k}+m^{2}}
\end{eqnarray}
The pair creation rate $w$ per $D$-dimensional volume $V_{D}$ is 
\begin{equation}
\label{pcrate}
w=-2 Im\,\cal{F}.
\end{equation}

In unoriented string theories, the one-loop diagrams of the same string coupling order (Euler number$=0$) are composed of the annulus $\cal{A}$, M\"{o}bius $\cal{M}$, the torus $\cal{T}$, and Klein bottle $\cal{K}$ \cite{angelaton02}, 
\begin{equation}
\label{oneloop}
\ln Z=\cal{A}+\cal{M}+\cal{T}+\cal{K},
\end{equation}
among which the annulus diagram $\cal{A}$ turns out to be the most relevant to the pair creation \cite{bachas92}. In type-IIA or type-IIB theories, being oriented theories, only  the annulus $\cal{A}$ and the torus $\cal{T}$ come into play. Here also, the former is relevant  because the closed strings do not couple to the gauge fields, thus do not produce the imaginary part of the vacuum energy density unless the closed string vacuum itself is unstable. 

Introducing Schwinger's proper time $t$, one can write the annulus diagram as 
\begin{eqnarray}
\label{annulus}
\cal{A} & = & \int^{\infty}_{0}\frac{dt}{2t}\,\text{Tr} e^{-2\pi t (H^{\|}+H^{\bot})}\nonumber \\
 & = &  \int^{\infty}_{0}\frac{dt}{2t}\,\text{Tr}_{\|} e^{-2\pi t H^{\|}}\text{Tr}_{\bot} e^{-2\pi t H^{\bot}}\nonumber \\
 &=&  \int^{\infty}_{0}\frac{dt}{2t}\,A^{\|}\cdot A^{\bot},
\end{eqnarray}
where the longitudinal part $A^{\|}$ (involved with the non-integer modes) and the transverse part $A^{\bot}$ (along the directions $\mu=2,\,3,\cdots,9$) have been factorized. Each part is further divided into the bosonic part and the fermionic part. 
The details concerning the full Hamiltonian are given in App. \ref{appii}.

\subsection{The Longitudinal Part}

Let us first specialize to the longitudinal part;
\begin{eqnarray}\label{parallel}
A^{\|}& = & \text{Tr}_{\|}e^{-2\pi t(H^{\|}_{B}+H^{\|}_{F})}\nonumber\\
 & = & \frac{|G|}{ 4\pi^{2}\alpha'}\int dx^{0}\,dx^{1}\, \text{tr}\,e^{-2\pi t H^{\|}_{B}}\cdot \text{tr}\,e^{-2\pi t H^{\|}_{F}}.
\end{eqnarray}
The front factor appears as a result of the normalization of the the `phase space' volume $[x^{0},\,x^{1}]=2\pi i \alpha'/|G|$. The precise expression of $G$ can be read from the corresponding commutator in Eq. (\ref{commutator}). The small traces `$\text{tr}$' are over the string states. The longitudinal part contains the reparametrization ghosts ($b, c$), and the superconformal ghosts ($\beta, \gamma$) in addition to the non-integer modes. The bosonic part is summarized as
\begin{eqnarray}\label{lboson}
 \text{tr}\,e^{-2\pi t H^{\|}_{B}}& = & \left(q^{{\cal E}^{\|}_{X}}\prod^{\infty}_{n=1}\frac{1}{1-q^{n-i\epsilon}}\prod^{\infty}_{m=0}\frac{1}{1-q^{m+i\epsilon}}\right)
  \cdot \left(q^{{\cal E}_{bc}}\prod^{\infty}_{l=1}(1-q^{l})^{2}\right) \nonumber\\
 &=&\frac{q^{{\cal E}^{\|}_{X}+{\cal E}_{bc}}}{1-q^{i\epsilon}}\prod^{\infty}_{n=1}\frac{1-q^{n}}{1-q^{n-i\epsilon}}\prod^{\infty}_{m=1}\frac{1-q^{m}}{1-q^{m+i\epsilon}}, 
\end{eqnarray}
where `$q$' represents `$\exp{(-2\pi t)}$'. In the sequel, we will see that the novel factor $1/(1-q^{i\epsilon})$, present only when $\epsilon\ne 0$, causes the pair creation. The zero point energies are 
\begin{eqnarray}
{\cal E}^{\|}_{X}+{\cal E}_{bc} & = & \left(\frac{i\epsilon}{2}(1-i\epsilon)-\frac{1}{12}\right)+\frac{1}{12}\\
 & = & \frac{i\epsilon}{2}(1-i\epsilon)\,. 
\end{eqnarray}

The longitudinal fermion part comes in a similar fashion;
\begin{eqnarray}
 \text{tr}\,e^{-2\pi t H^{\|}_{F}}& = & \left(q^{{\cal E}^{\|}_{\psi}}\prod^{\infty}_{r>0}(1+q^{r-i\epsilon})\prod^{\infty}_{s\geq 0}(1+q^{s+i\epsilon})\right)\cdot\left(\frac{q^{{\cal E}_{\beta\gamma}}}{m}\prod^{\infty}_{u>0}\frac{1}{(1+q^{u})^{2}}\right) \\
 & = & q^{{\cal E}^{\|}_{\psi}+{\cal E}_{\beta\gamma}}{\cal M}\prod^{\infty}_{r>0}\frac{(1+q^{r-i\epsilon})(1+q^{r+i\epsilon})}{(1+q^{u})^{2}}, 
\end{eqnarray}
where the indices, $r,\,s,\,u$ are half-integer valued for NS-sector and integer valued for R-sector. The multiplicity `$m$' (of the ground state), in the first line, is `$1$' for NS-sector, but for R-sector it is `$2$' due to the ghost zero mode $\gamma_{0}$ (of ($\beta,\,\gamma$)). In the same vein, the factor $\cal{M}$ is `$1$' for NS-sector and $(1+q^{i\epsilon})/2$ for R sector. The zero point energies are
\begin{eqnarray}
{\cal E}^{\|}_{\psi}+{\cal E}_{\beta\gamma} & = & 
\left\{
\begin{array}{ll}
\displaystyle\left(-\frac{i\epsilon}{2}(1-i\epsilon)+\frac{1}{12}\right)-\frac{1}{12}= -\frac{i\epsilon}{2}(1-i\epsilon) & \quad\text{(R)}\\
{}&{}\\
\displaystyle\left(\frac{(i\epsilon)^{2}}{2}-\frac{1}{24}\right)+\frac{1}{24} =\frac{(i\epsilon)^{2}}{2} & \quad\text{(NS)}
\end{array}\right.
\end{eqnarray}

\subsection{The Transverse Part}

Let us denote the transverse directions as $\bot=\{2,\,3,\,\cdots, 9\}$. The factor $A^{\bot}$ in Eq. (\ref{annulus}) is
\begin{eqnarray}\label{anormal}
A^{\bot} & = & \text{Tr}_{\bot}e^{-2\pi t(H^{\bot}_{B}+H^{\bot}_{F})} \nonumber\\
 & = & \int dx^{2}\int \frac{dk^{2}}{2\pi}\,\text{tr}\,e^{-2\pi t H^{\bot}_{B}}\cdot \text{tr}\,e^{-2\pi t H^{\bot}_{F}}.
\end{eqnarray}
More specifically, the bosonic part is
\begin{eqnarray}\label{normalboson}
\displaystyle \text{tr}\,e^{-2\pi t H^{\bot}_{B}}& = & q^{\frac{1}{2}\alpha^{2}_{0}+\frac{y^{2}}{4\pi^{2}\alpha'}-\frac{1}{3}}\prod^{\infty}_{n=1}\frac{1}{(1-q^{n})^{8}}\,. 
\end{eqnarray}
Here we note that the term $y^{2}/(4\pi^{2}\alpha')$ in the exponent is due to the separation between D-branes, and the zero point energy for the transverse bosonic matter part is ${\cal E}^{\bot}_{X}=(-1/24)\times 8=-1/3$.

Lastly, the fermionic part is
\begin{equation}
\label{tranfermion}
\text{tr}\,e^{-2\pi tH^{\bot}_{F}}=q^{{\cal E}^{\bot}_{\psi}}\prod^{\infty}_{r>0}\left(1+q^{r}\right)^{8},
\end{equation}
where the transverse fermion zero point energy is
\begin{eqnarray}
{\cal E}^{\bot}_{\psi}& = & -\frac{8}{2}\left(\frac{1}{24}-\frac{1}{8}(2\vartheta-1)^{2}\right) \nonumber\\
 & = & \left\{\begin{array}{rl}
    \displaystyle \frac{1}{3}  &  \quad\text{(R;$\quad\vartheta=0)$} \\
    {}&{}\\
    \displaystyle -\frac{1}{6}&  \quad\text{(NS;\,\,$\vartheta=\frac{1}{2}$)} 
\end{array}\right. 
\end{eqnarray}

\subsection{D-D Pair vs. D-$\overline{\text{D}}$ Pair}

The notion of D-D pair and D-$\overline{\text{D}}$ pair depends on the way of GSO projection. As for NS-sector and R-sector, GSO projection is performed as
\begin{eqnarray}
\text{tr}\, e^{-2\pi t H_{\text{NS,R}}}& \Rightarrow &\text{tr}  \left(\frac{1+\lambda e^{\pi i F}}{2}\right) e^{-2\pi t H_{\text{NS,R}}},
\end{eqnarray}
where $F$ is the world-sheet fermion number defined mod $2$ so that $e^{\pi i F}$ anticommute with all the fermion modes $\varphi^i_r$ ($i=2,3,\cdots,9$). Especially on the R-sector ground states, the operator $e^{\pi i F}$ is represented as the chiral operator of $SO(8)$ Clifford algebra composed of $8$ zero modes $\varphi^i_0$. The value $\lambda=1$ for D-D pair and $\lambda=-1$ for D-$\overline{\text{D}}$ pair.
Putting it altogether, one can write the result as
\begin{eqnarray}\label{annulus1}
{\cal A} & = & \frac{|G\,\xi|V_3}{2\alpha'}\int^{\infty}_{0}\frac{dt\, \,q^{\frac{y^2}{4\pi^2\alpha'}}}{(2\pi)^3 t^{3/2}}\left[\frac{
q^{\frac{i\epsilon}{2}-\frac{1}{2}}
}
{1-q^{i\epsilon}}
\left(
\prod^{\infty}_{n=1}\frac{(1+q^{n-\frac{1}{2}-i\epsilon})(1+q^{n-\frac{1}{2}+i\epsilon})(1+q^{n-\frac{1}{2}})^6}{(1-q^{n-i\epsilon})(1-q^{n+i\epsilon})(1-q^n)^6}\right.\right.\nonumber\\
&&\qquad\left.-\lambda
\prod^{\infty}_{n=1}\frac{(1-q^{n-\frac{1}{2}-i\epsilon})(1-q^{n-\frac{1}{2}+i\epsilon})(1-q^{n-\frac{1}{2}})^6}{(1-q^{n-i\epsilon})(1-q^{n+i\epsilon})(1-q^n)^6}
\right)\nonumber\\
&&\qquad\left.-8\frac{1+q^{i\epsilon}}{1-q^{i\epsilon}}\prod^{\infty}_{n=1}
\frac{(1+q^{n-i\epsilon})(1+q^{n+i\epsilon})(1+q^{n})^6}{(1-q^{n-i\epsilon})(1-q^{n+i\epsilon})(1-q^{n})^6}\right] \nonumber\\
&=&\frac{i|G\,\xi|V_3}{2\alpha'}\int^{\infty}_{0}\frac{dt\, \,q^{\frac{y^2}{4\pi^2\alpha'}}}{(2\pi)^3 t^{3/2}}\frac{1}{
\theta_1(\epsilon t|it)\,(\eta(it))^{9}}\times\nonumber\\
&&\qquad\times\left[\theta_3(\epsilon t|it)\,\theta^3_3(0|it)-\lambda\,\theta_4(\epsilon t|it)\,\theta^3_4(0|it)-\theta_2(\epsilon t|it)\,\theta^3_2(0|it)
\right]\,.
\end{eqnarray}
A few explanations are in order. The front factor $|\xi|$ and the additional fractional part $1/2$ in the index `$3/2$' of the proper time $t$ are due to the integration of the factor $q^{\alpha^2_0/2}$ (in Eq. (\ref{normalboson})) over the momentum component $k^2$ (see Eq. ( \ref{momentum})). Had we worked with higher dimensional Dp-branes, we would have obtained higher power for $\xi$ and $t$. Indeed the directions normal to the vector $\vec{e}-\vec{e'}$ will be $(p-1)$-dimensional and the integrations in Eq. (\ref{anormal}) will be extended as such, 
\begin{eqnarray}
\int\frac{d^{p-1}\vec{k}}{(2\pi)^{p-1}}\,e^{-\frac{\pi t}{\xi^2}(\vec{k})^2}  =  \frac{|\xi|^{p-1}}{(2\pi)^{p-1}t^{\frac{p-1}{2}}}\,\,. 
\end{eqnarray}
 
The common factor $(1-q^{i\epsilon})$ in the denominators in Eq. (\ref{annulus1}) originates from the zero mode of the longitudinal bosonic matter part $X^\|$ (see Eq. (\ref{lboson})), while the factor $(1+q^{i\epsilon})$ in the numerator of the R-sector (the third term) is due to the zero mode of the longitudinal fermionic matter part $\psi^\|$. We summed over all the zero point energies, which makes $(i\epsilon -1)/2$ for the NS sector. The whole R-sector does not have zero point energy, as it should be.   
In the last line, we recast the terms by making use of theta function;
\begin{eqnarray}\label{thetaftn}
&&\theta[{}^{\alpha}_{\beta}](z|\tau)  =  e^{2\pi i \alpha(z+\beta)}q^{\alpha^2/2}\prod^{\infty}_{n=1}(1-q^{n})(1+q^{n+\alpha-\frac{1}{2}}e^{2\pi i (z+\beta)})(1+q^{n-\alpha-\frac{1}{2}}e^{-2\pi i(z+\beta)})\,,\nonumber\\
&&\\
&&\theta_1(z|\tau)=\theta[{}^{1/2}_{1/2}](z|\tau),\quad\theta_2(z|\tau)=\theta[{}^{1/2}_{0}](z|\tau),\quad\theta_3(z|\tau)=\theta[{}^{0}_{0}](z|\tau),\quad\theta_4(z|\tau)=\theta[{}^{0}_{1/2}](z|\tau).\nonumber
\end{eqnarray}

We note that the integrand vanishes if $\epsilon\rightarrow 0$ and $\lambda=1$ due to Jacobi's `abstruse identity'. Therefore, when the background fields are such that $\beta^{2}\rightarrow 0$ (See Eq. (\ref{beta})), the contribution from bosons and fermions exactly cancel each other for D-D pair ($\lambda=1$). This is just one of the BPS configurations discussed in detail in Ref \cite{cho03}.  

From Eq. (\ref{pcrate}), we note that pair creations occur when ${\cal A}$ is  real valued. In the case at hand, the integrand is purely imaginary (taking complex conjugate results in over all sign change), there seems no pair creation. However, the common factor $1/(1-q^{i\epsilon})$ makes simple poles of the integrand at $t=l/\epsilon$ ($l=1,\,2,\,3,\cdots$) \footnote{In fact, the higher order pole at $k=0$ makes the integration divergent. This could be remedied by adjusting the graviton and dilaton background. In this paper, we rather focus on the simple poles which are relevant to the pair creation.}. (Note that we assumed $\epsilon$ to be positive in an earlier section.) The contour integration will give the blessed imaginary factor `$\pi i$', thus the annulus becomes real valued. Hence, technically we note that the main contribution to the pair creation comes from the zero mode of the longitudinal bosons which generates the singular factor $1/(1-q^{i\epsilon})$ (Note that this factor is absent when $\epsilon\rightarrow 0$, as is clear from Eq. (\ref{lboson})).

\section{Open String Pair Creation Rate}\label{secv}

\subsection{On Shell Conditions}

After the contour integration (with the contour passing over the positive real axis in complexified $t$-plane), we are left with only the transverse part. 

\begin{eqnarray}\label{partitionrslt}
{\cal A} & = & \frac{|G\,\xi|V_3}{(2\pi)^3\alpha'}\sum^{\infty}_{l=1}\left(\frac{\epsilon}
{l}\right)^{\frac{3}{2}}e^{-\frac{y^{2}l}{2\pi\alpha'\epsilon}}\frac{1}{2\epsilon}\times\\
&&\times\left.\left[\frac{(-1)^l}{2}q^{-\frac{1}{2}}\left(\prod^{\infty}_{n=1}\frac{(1+q^{n-1/2})^8}{(1-q^n)^8}-\lambda\prod^{\infty}_{n=1}\frac{(1-q^{n-1/2})^8}{(1-q^n)^8}\right)-8\prod^{\infty}_{n=1}\frac{(1+q^{n})^8}{(1-q^n)^8}\right]\right\vert_{q=e^{-2\pi l/\epsilon}}\,\,\nonumber\\
&=&\frac{|G\,\xi|V_3}{(2\pi)^3\alpha'}\sum^{\infty}_{l=1}\left(\frac{\epsilon}
{l}\right)^{\frac{3}{2}}e^{-\frac{y^{2}l}{2\pi\alpha'\epsilon}}\frac{1}{4\epsilon}
\frac{1}{\eta^{12}(\frac{il}{\epsilon})}\displaystyle\Big[(-1)^l\left(1-\lambda\right)\theta^4_4(0\vert\textstyle{\frac{il}{\epsilon}})+
\left((-1)^l-1\right)\theta^4_2(0\vert \textstyle{\frac{il}{\epsilon}})\Big]\,.\nonumber
\end{eqnarray}
In the last line, we used Jacobi's `abstruse identity';
\begin{equation}
\label{jacobiid}
\theta^4_2(0|q)-\theta^4_3(0|q)+\theta^4_4(0|q)=0.
\end{equation}

We note that each of these terms, coming from the trace
\begin{eqnarray}
\left.\text{tr} \frac{\left(1+\lambda e^{i\pi F}\right)}{2}e^{-2\pi t H^\bot}\right|_{t=l/\epsilon},
\end{eqnarray}
can be cast into a simpler form in terms of the mass spectrum.
On shell, the exponent vanishes;
\begin{eqnarray}\label{massop}
H^\bot&=&\frac{(\hat{k}^2)^2}{2\xi^2}+\frac{y^2}{4\pi^2\alpha'}+{\cal E}_0+\sum^9_{i=2}\left(\sum^{\infty}_{n=1}n\hat{N}^i_n+\sum_{r} r\hat{N}^i_r\right)\nonumber\\
&\equiv&\frac{1}{2\xi^2}\left((\hat{k}^2)^2+\frac{\xi^2y^2}{2\pi^2\alpha'}+\hat{M}^2\right)=0
\end{eqnarray} 
The last line defines the mass operator $\hat{M}$ contributed by the oscillatory part. Although the inter-string, being stretched between D-branes, has extra contribution $\frac{y^2}{4\pi^2\alpha'}$ to mass square, we leave it aside deliberately to see its effect on the pair creation rate. 
The subindex $r$ takes integer (R sector) or half integer values (NS sector). In the first line, ${\cal E}_0$ is the total zero point energy whose value is
\begin{eqnarray}
{\cal E}_0 & = & {\cal E}^{\|}_{X}+{\cal E}_{bc}+{\cal E}^{\|}_{\psi}+{\cal E}_{\beta\gamma}+{\cal E}^{\bot}_{X}+{\cal E}^{\bot}_{\psi}\nonumber\\
 & = & \left\{\begin{array}{ll}
    \displaystyle 0  &  \quad\text{(R)} \\
    {}&{}\\
    \displaystyle -\frac{1}{2}(1-i\epsilon)&  \quad\text{(NS)} 
\end{array}\right. 
\end{eqnarray} 
Therefore for most on-shell states, the spatial momentum component $k^2$ is complex valued, and so is the temporal component, $k^0$, via the relation (\ref{in}). Although this looks strange, we will see below that it is rather the ratio of these components that is physically sensible.

We finally obtain
\begin{equation}
\label{final}
{\cal A}=-\frac{|G\,\xi|V_3}{(2\pi)^3\alpha'}\sum^{\infty}_{l=1}\left(\frac{\epsilon}
{l}\right)^{\frac{3}{2}}e^{-\frac{y^{2}l}{2\pi\alpha'\epsilon}}\frac{1}{2\epsilon}\left.\left((-1)^{l+1}\sum_{S\in \text{\{NS\}}}q^{\frac{M^2_{S}}{2\xi^2}}+\sum_{S\in \text{\{R\}}}q^{\frac{M^2_{S}}{2\xi^2}}\right)\right\vert_{q=e^{-2\pi l/\epsilon}}\,.
\end{equation} 
Although this expression applies to both values of $\lambda=\pm 1$, the summations over states $S$ distinguishes D-D and D-$\overline{\text{D}}$ cases. For example, the NS ground state with $M^2=\xi^2(i\epsilon-1)$ survives GSO projection for D-$\overline{\text{D}}$ case, while it is absent in D-D case. The exponential factors inside the big parentheses are universal ones in the pair creation phenomena and can be also obtained by minimizing the Euclidean effective action for the inter-string \cite{gorsky01}

Let us look at the asymptotic behavior of $\cal{A}$ in more detail when  $\epsilon\ll 1$.
For each $l$, we take $q=\exp{(-2\pi l/\epsilon)}\ll 1$ as an expansion parameter and compare the order difference between D-D case and D-$\overline{\text{D}}$ case. As for D-D case ($\lambda=1$), only the second term involving $\theta^{4}_{2}$ in the last line of Eq. (\ref{partitionrslt}) survives. Meanwhile for D-$\overline{\text{D}}$ case ($\lambda=-1$), the identity (\ref{jacobiid}) simplifies the expression so that only the terms involving $\theta^{4}_{3}$ and $\theta^{4}_{4}$ remain. The basic order difference of these theta functions can be read from their defining equation, (\ref{thetaftn}); 
\begin{eqnarray}
&&{\cal A}=\frac{\epsilon^{\frac{1}{2}}|G\,\xi|V_3}{4(2\pi)^3\alpha'}\sum^{\infty}_{l=1}\left(\frac{1}
{l}\right)^{\frac{3}{2}}e^{-\frac{y^{2}l}{2\pi\alpha'\epsilon}}\times\\
&&\times\left\{\begin{array}{cc}\displaystyle\left((-1)^l-1\right)\frac{\theta^4_2(0\vert \textstyle{\frac{il}{\epsilon}})}{\eta^{12}(\frac{il}{\epsilon})}\quad\sim\quad \displaystyle16\left((-1)^l-1\right)(1+{\cal O} (e^{-\frac{2\pi l}{\epsilon}}))&\quad\text{(D-D)} \\
&\nonumber\\
\displaystyle\left((-1)^l-1\right)\frac{\theta^4_3(0\vert \textstyle{\frac{il}{\epsilon}})}{\eta^{12}(\frac{il}{\epsilon})}+\left((-1)^l+1\right)\frac{\theta^4_4(0\vert \textstyle{\frac{il}{\epsilon}})}{\eta^{12}(\frac{il}{\epsilon})}\quad\sim\quad \displaystyle 2\, e^{\frac{\pi l}{\epsilon}}(-1)^{l}(1+{\cal O}(e^{-\frac{\pi l}{\epsilon}}))&\quad\text{(D-$\overline{\text{D}}$)}\end{array}\right.
\end{eqnarray}
When the separation between branes is larger than sub-stringy order, more specifically $y >\sqrt{2\pi^{2}\alpha'}$, the contributions from higher values of $l$ are exponentially suppressed for both cases. Therefore the leading order ($l=1$) shows that the string pair creation on D-D branes, is more suppressed by the factor $\exp{(-\pi/\epsilon)}$ in comparison with D-$\overline{\text{D}}$ case. On the other hand for the substringy separation of D-$\overline{\text{D}}$ branes, that is when  $y<\sqrt{2\pi^{2}\alpha'}$, higher values of $l$ contribute more but with alternating sign $(-1)^{l}$. The critical value of the separation $y=\sqrt{2\pi^{2}\alpha'}$ is just the scale where the  GSO-projected ground state of the inter-string becomes massless with background fields turned off (see Eq. (\ref{massop})). It would be interesting to see the sub-stringy physics, which will be pursued elsewhere.

\subsection{Higher Dimensional Extension}

The pair creation rate per volume (as defined in Eq. (\ref{pcrate})) is
\begin{eqnarray}\label{rate}
\omega & = & \frac{|G\,\xi|}{(2\pi)^3\alpha'}\left.\sum_{S\in \text{\{NS,R\}}}\sum^{\infty}_{l=1}\left(\frac{\epsilon}
{l}\right)^{\frac{3}{2}}\frac{e^{-\frac{y^{2}l}{2\pi\alpha'\epsilon}}}{\epsilon}(-1)^{(l+1)(a_S+1)}q^{\frac{M^2_{S}}{2\xi^2}}\right\vert_{q=e^{-2\pi l/\epsilon}}\,,
\end{eqnarray}
where $a_{\text NS}=0$ and $a_{\text R}=1$.
As for the case of a D$p$-brane pair, the front factor and the power `$3/2$' of $\epsilon$ are modified as
\begin{eqnarray}
 \frac{|G\,\xi|}{(2\pi)^3\alpha'} &\rightarrow & \frac{|G\,\xi^{p-1}|}{(2\pi)^{p+1}\alpha'}, \\
\left(\frac{\epsilon}
{l}\right)^{\frac{3}{2}} & \rightarrow & \left(\frac{\epsilon}
{l}\right)^{\frac{p+1}{2}}.
\end{eqnarray}

In the above expressions and in Eq. (\ref{final}), we note that the factor $G$ is not symmetric under the exchange $(\vec{e},\,\vec{b})\leftrightarrow(\vec{e'},\,\vec{b'})$ (see the commutator $[x^{0},\,x^{1}]=-2\pi i \alpha'/|G|$ in Eq. (\ref{commutator})). This is a consequence of the orientation of the string. Actually, we should have summed over the orientation too. The change will be achieved by rewriting the factor $G$ in the symmetrized form;
\begin{eqnarray}
&&\frac{|G\,\xi^{p-1}|}{(2\pi)^{p+1}\alpha'} =  \frac{|\vec{e}-\vec{e'}|^{p-1}}{(2\pi)^{p+1}(2\alpha')^{\frac{p+1}{2}}}\left(-\beta^2(1-\vec{e}\cdot\vec{e'}+\vec{b}\cdot\vec{b'})^2\right)^{\frac{3-p}{2}}\times\\
&&\times\left[\frac{1}{\left|e'_1(1-e_2e'_2+\vec{b}\cdot\vec{b'})-e_1(1-e_2e'_2+\vec{b'}\cdot\vec{b'})\right|}+\frac{1}{\left|e_1(1-e_2e'_2+\vec{b}\cdot\vec{b'})-e'_1(1-e_2e'_2+\vec{b}\cdot\vec{b})\right|}\right] .\nonumber
\end{eqnarray}

\section{Various Dual Configurations}\label{secvi}

\subsection{String Unit}

Let us specify the string unit in the presence of background fields. The point in defining the string unit is to equate the momentum $k^{\mu}k_\mu$ to the string zero mode $(\alpha_{0})^2$. From Eq. (\ref{spacelike}), we note that
the string unit can be defined by setting
\begin{equation}
\label{stringunit }
\alpha'=-\frac{|\vec{e}-\vec{e'}|^{2}-|\vec{b}-\vec{b'}|^{2}}{2\beta^{2}(1-\vec{e}\cdot\vec{e'}+\vec{b}\cdot\vec{b'})^{2}}
\end{equation} 
In other words, if we define the effective string tension as
\begin{eqnarray}
T_{\text{eff}} & = & \frac{1}{2\pi\alpha'_{\text{eff}}}=-\frac{|\vec{e}-\vec{e'}|^{2}-|\vec{b}-\vec{b'}|^{2}}{2\pi\alpha'\beta^{2}(1-\vec{e}\cdot\vec{e'}+\vec{b}\cdot\vec{b'})^{2}} \\
 & = & -\frac{|\vec{e}-\vec{e'}|^{2}-|\vec{b}-\vec{b'}|^{2}}{\beta^{2}(1-\vec{e}\cdot\vec{e'}+\vec{b}\cdot\vec{b'})^{2}}\, T,
\end{eqnarray}
the above string unit sets $2\alpha'_{\text{eff}}=1$.
The factor $\xi$ defined in Eq. (\ref{momentum}) is then
\begin{equation}
\label{factor }
\frac{1}{\xi^{2}}=\frac{|\vec{e}-\vec{e'}|^{2}-|\vec{b}-\vec{b'}|^{2}}{|\vec{e}-\vec{e'}|^{2}}.
\end{equation}

\subsection{Comparison with Schwinger's Case and Bachas-Porrati's Case}

Let us specialize to the case discussed in Ref. \cite{bachas92}. As was already noted in Eq. (\ref{bachas}), it corresponds to the case of $e_{1}=-B^{(0)}_{01}=-\pi q_{1}E$ and $e'_{1}=-B^{(\pi)}_{01}=\pi q_{2}E$, that is, the electric fields, $\vec{e}$ and $\vec{e'}$, are parallel or anti-parallel depending on the charges $q_{1}$ and $q_{2}$. There is no magnetic field, so $\vec{b}=\vec{b'}=0$.\footnote{One might be tempted to take a simple boosting to achieve this specific case. This is possible as will be discussed in later part of Sec. \ref{6.5.2}. However, we have to note that the boosting procedure can suppress the magnetic fields almost but not completely. Actually the reverse procedure is not possible, that is, one cannot recover different values of magnetic fields from the case with the electric fields only.} The string unit in this case takes the ordinary form $\alpha'=1/2$ and $\xi^{2}=1$. From Eq. (\ref{epsilon}), we note that 
\begin{equation}
\label{tanhbachas}
\tanh^{2}(\epsilon\pi)=\frac{|\vec{e}-\vec{e'}|^{2}}{(1-\vec{e}\cdot\vec{e'})^{2}}.
\end{equation}
Therefore, $\epsilon\pi=-\tanh^{-1}e_{1}+\tanh^{-1}e'_{1}=\tanh^{-1}(\pi q_{1}E)+\tanh^{-1}(\pi q_{2}E)$.
The open string pair creation rate, for the case of D$p$-(anti-)D$p$ branes, reduces (in the string unit) to
\begin{eqnarray}\label{bachaslimit}
\omega & = & \frac{2|\vec{e}-\vec{e'}|}{(2\pi)^{p+1}}\sum_{S\in \text{\{NS,R\}}}\sum^{\infty}_{l=1}\left(\frac{\epsilon}
{l}\right)^{\frac{p+1}{2}}\frac{e^{-\frac{y^{2}l}{\pi\epsilon}}}{\epsilon}(-1)^{(l+1)(a_S+1)}e^{-\frac{\pi k M^2_{S}}{\epsilon}},
\end{eqnarray}
which is in accordance with the result of \cite{bachas92} except overall factor `2'. This is due to the fact that the spectrum of the unoriented string is halved by the orientation projection, compared to that of IIA or IIB string. The reason why the result (\ref{annulus1}) takes the same form as those of Refs. \cite{bachas92} and \cite{bachas95} is that there are only a few Lorentz- and gauge- invariant combinations of gauge fields with which one can construct the novel imaginary mode in Eqs. (\ref{beta}) and (\ref{epsilon}). They are $B^{(0)}_{\mu\nu }B^{(0)\mu\nu }$, $B^{(0)}_{\mu\nu }B^{(\pi)\mu\nu }$, and $B^{(\pi)}_{\mu\nu }B^{(\pi)\mu\nu }$.

In the weak field limit \footnote{Recall that $B_{\mu\nu}$ represents the demensionless quantity `$2\pi\alpha'$ times the field strength'. The weak field limit could be considered as $\alpha'\rightarrow 0$ limit.}, 
\begin{equation}
\label{weakfield}
\epsilon\pi=(e'_{1}-e_{1})+\frac{1}{3}((e'_{1})^{3}-(e_{1})^{3})+{\cal O}(e^{5}, e'^{5}),
\end{equation}
with the insertion of the lowest order of which into the above equation (\ref{bachaslimit}), one recovers Schwinger's original 
result in Ref. \cite{schwinger51}.

\subsection{a Hanany-Witten like Effect? (dual to the case of $\vec{e}\bot\vec{e'}$, $\vec{b}=\vec{b'}=0$)}\label{orthogonal}

\FIGURE{
\epsfbox{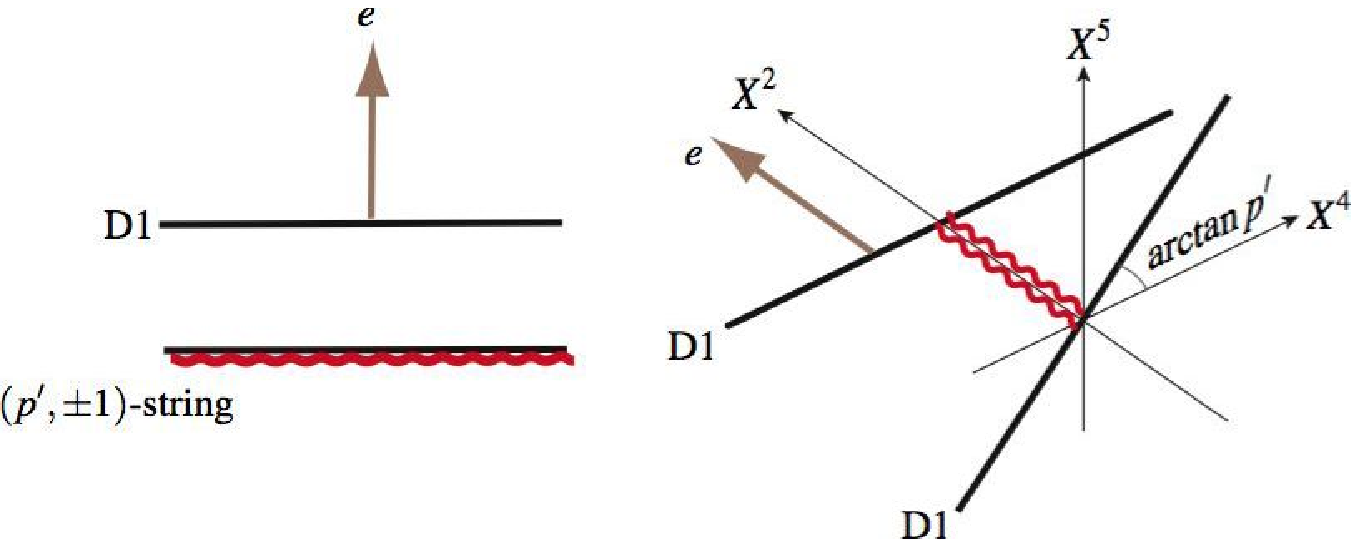}\label{fig2}
\caption{The left figure shows IIB configurations T-dual to a D$2$-(anti-)D$2$ pair with $\vec{e}\bot\vec{e'}$, $\vec{b}=\vec{b'}=0$. T-duality has been taken along the direction of $\vec{e}$. The (anti-)D$2$-brane with the electric field $\vec{e'}$ orthogonal to the direction of $\vec{e}$ becomes a $(p',\,\pm1)$-string at rest. The number $p'$ of F-strings is determined by the electric displacement of the field $\vec{e'}$. The right IIB configuration obtained through a series of duality transformations on the left configuration shows a feature similar to (not exactly the same as) Hanany-Witten effect: two D-strings `passed through' by each other generate a pair of strings connecting them. In fact, they are separated in the residual transverse dimensions by a distance $y$.}
}

Let us look around other examples of configurations involving pair creations. 
The first nontrivial example is the case where the electric vectors $\vec{e},\,\vec{e'}$ are orthogonal to each other. To simplify the argument, we let $e=e'$, $\vec{b}=\vec{b'}=0$ and $\theta=\theta'=\pi/4$. One can easily see that the criterion for the pair creation is satisfied; $\beta^{2}=-e^{2}(1+(1-e^{2}))<0$. In the string unit of this very case, $\alpha'=1/(2-e^{2})$ and the front factor of the pair creation rate is
\begin{equation}
\label{normal}
\frac{|G\xi|}{\alpha'}=\sqrt{2}\,e\,(2-e^{2}).
\end{equation}
We observe that unlike previous case, this has cubic dependence on $e$ in the front factor of the pair creation rate (\ref{rate}).

It is amusing to expect open string pair creation in the T-dual configurations. Under the T-duality along the vector $\vec{e}$ (thus, normal to $\vec{e'}$), the D$2$-brane becomes a D-string moving with the velocity $e$ and the (anti-)D$2$-brane becomes a BPS bound state of an (anti-)D-string and $p'$ F-strings at rest (where the number $p'$ is determined by the electric displacement of the field $\vec{e'}$). The distance $y$ between the D$2$- and (anti-)D$2$-brane of IIA setup corresponds to the impact parameter for the D-strings passing by each other in IIB theory. Therefore the physics in this dual setup is the string pair creation in the system of moving D-strings and standing still $(p',\,\pm 1)$-strings. See the left figure in Fig. 2.

Taking more dualities including S-duality, one can achieve more interesting configuration. To be more specific, we start from the IIB picture, the left configuration in Fig. 2. Let the static $(p',\pm 1)$-string laid along $X^1$-direction and the D-string move along $X^2$-direction. After a journey of various dualities, $T_5\circ T_1\circ S\circ T_4\circ T_5$, we get to two D-strings; one is along $X^4$-direction and moves along $X^2$-axis, and the other is slanted with tilting angle, $\arctan{p'}$, with respect to $X^3$-axis in $(3,4)$-plane, and is static but `passed through' by the first D-string (the right figure in Fig. 2). The string pair creation in the original IIA configuration corresponds to the pair creation of strings between two D-strings at an angle when one passes the other. 

Although this latter phenomenon looks very similar to Hanany-Witten effect \cite{Hanany:1997}, there are some differences. First, those two D-strings are {\it not actually} passed through by each other, but are separated by a distance $y$ in the residual transverse directions. The pair creation is exponentially suppressed as the distance $y$ increases. Another difference is that the strings are created in pair, featuring the non-supersymmetric nature of the configuration \cite{Kitao:1998}. This is in contrast with Hanany-Witten effect that is related with the anomaly inflow mechanism via U-duality \cite{Bachas1997}. In the anomaly inflow mechanism, the presence of unbroken chiral supersymmetries is essential (see Ref. \cite{Green1997} for details).

\FIGURE{
\epsfbox{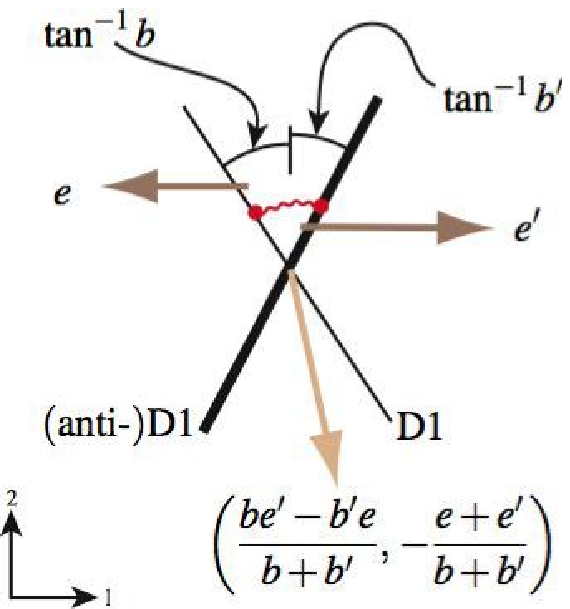}\label{hanany}
\caption{The configuration is T-dual to the case of a D$2$-(anti-)D$2$ pair with anti-parallel electric fields and anti-parallel magnetic fields. As two D-strings move to the opposite directions, their intersection moves with the velocity $(be'-b'e,\,\,-e-e')/(b+b')$.}
}

\subsection{Spacelike Scissors (dual to the case of $\vec{e}=-\vec{e'}\ne0$, $\vec{b}=-\vec{b'}\ne0$)}

In order to see how the magnetic field affects the pair creation, let us add magnetic fields $\vec{b}$ and $\vec{b'}$ to the case of $\vec{e}+\vec{e'}=0$. For simplicity, we let $\vec{b}+\vec{b'}=0$. Since 
\begin{equation}
\beta^{2}=\frac{4(b-e)(b+e)}{(1-b^{2}+e^{2})^{2}},
\end{equation}
the criterion of $\beta^{2}<0$ is satisfied only when $b<e$. Hence, it is impossible to consider the pair creation in the purely magnetic case. In the string unit ($\alpha'=1/2$), the front factor of the pair creation rate is
\begin{equation}
\label{parallel1}
\frac{|G\xi|}{\alpha'}=4\sqrt{e^{2}-b^{2}}.
\end{equation}

The role of the magnetic fields becomes more transparent in the T-dual setup. T-duality along the direction of the electric fields leads us to two tilted D-strings
intersecting at an angle $\pi-2\arctan{b}$, and moving with the velocity $\vec{e}$ and $-\vec{e}$ respectively. One interesting thing is that the intersection moves along $X^{2}$-direction with the speed $e/b$. Therefore, the criterion for the pair creation is obeyed when the intersection moves at the super-luminal speed. This is the spacelike scissors discussed in Ref. \cite{bachas02} as an example of unstable configurations. (We will discuss its consequence in the next subsection.) In all, the magnetic fields regulate the tilting angle so that the pair creation ceases when $b$ approaches $e$, making null scissors \cite{bachas02,myers02,cho02,Chen2003}. (See also Refs. \cite{cho01,Lunin:2001} for more examples of BPS configurations of moving D-branes.) 

\subsection{Scissors Paradox and Its Resolution}
\subsubsection{the inter-string keeps up with the intersection}
As was mentioned previously, the configurations making pair creations show some bizarre behavior: the spacelike kinetic momentum (\ref{spacelike}) of the inter-string or the spacelike motion of the intersection of D-strings in IIB picture. Below, we will show that they are actually the same, that is, the inter-string in IIB setup catches up with the super-luminal intersection, thus makes the scissors paradox \cite{Taylor:1992}. We next propose a resolution based on the triple junction \cite{rey97}.

Since the string couples to the background gauge fields, the physical meaning of the spacelikeness of only the kinetic part of the full momentum (\ref{fullmomentum}) looks obscure. However, it really implies the spacelike motion of the inter-string. 
One can envisage the meaning of the kinetic momentum (\ref{momentumvec}) as follows. Since the oscillatory parts of the kinetic momentum and the field momentum cancel with each other, we write
\begin{equation}
\label{velocity}
\frac{1}{2\pi\alpha'}\int^\pi_0d\sigma\,\,\partial_\tau X^\mu=\frac{1}{2\alpha'}\frac{d<X^\mu>}{d\tau}\sim k^\mu.
\end{equation}
By `$\sim$' we mean, the equality up to the oscillatory part. Therefore the ratio of its components represents the velocity of the inter-string in space-time; 
\begin{equation}
\label{velocity2}
\frac{d <X^2>}{d <X^0>} \sim\frac{k^2}{k^0}=-\frac{|\vec{e'}-\vec{e}|}{|\vec{b'}-\vec{b}|}.
\end{equation}
Therefore the spacelikeness of the kinetic momentum, implying $|\vec{e'}-\vec{e}|>|\vec{b'}-\vec{b}|$ as in (\ref{spacelike}), leads to the super-luminal speed of the inter-string in space-time. 

Next, we obtain the velocity of the inter-string in a T-dual configuration. To be specific, let us take the electric fields to be anti-parallel along $X^1$-direction and the magnetic fields to be anti-parallel too. Since the constraint in (\ref{in}) is invariant under T-duality, as is shown in appendix C, the velocity component of the inter-string along $X^2$-direction is intact under T-duality taken along $X^1$-direction. In IIB configuration, the inter-string has the velocity component along $X^{1}$-direction too. Indeed the `winding mode',
\begin{eqnarray}
\triangle X^\mu\equiv X_\mu(\pi)-X_\mu(0)&\sim& 2\pi B_{\mu\nu}\,a^\nu_0\nonumber \\
& = & 2\pi a^2_0\left(\begin{array}{c}-e\cos\theta \\\\ \displaystyle \frac{b'e\sin\theta-be'\sin\theta'}{e\sin\theta+e'\sin\theta'} \\\\\displaystyle -\frac{(b+b') e\cos\theta}{e\sin\theta+e'\sin\theta'}\end{array}\right). 
\end{eqnarray} 
in the original theory becomes corresponding momentum component in the T-dual theory via the relation;
\begin{equation}
\frac{\triangle X^1}{2\pi\alpha'}\sim \frac{a^2_0}{\alpha'}\frac{(eb'-e'b)}{(e+e')}\rightarrow \tilde{k}^1.
\end{equation}
(Note that other components are trivial because $\theta=\theta'=\pi/2$.)
As before, the ratio $\tilde{k}^1/\tilde{k^0}=(be'-b'e)/(b+b')$ represents the velocity component of the inter-string along $\tilde{X}^1$-direction. Therefore, T-dual counterpart of the inter-string of IIA configuration moves at the velocity 
\begin{equation}
\label{intervelo}
\left(\frac{be'-b'e}{b+b'},\,\,-\frac{e+e'}{b+b'}\right).
\end{equation}

Lastly, we obtain the velocity of the intersection of D-strings. According to the prescription (\ref{bciib0}), the boundary condition in IIB configuration can be specified as
\begin{eqnarray}
\left.\partial_\sigma\left(\tilde{X}^0+B^{(\sigma)}_{01}\tilde{X}^1\right)\right|_{\sigma=0,\pi} & = & 0 \nonumber\\
\left.\partial_\tau\left(\tilde{X}^1-B^{(\sigma)}_{10}\tilde{X}^0-B^{(\sigma)}_{12}\tilde{X}^2\right)\right|_{\sigma=0,\pi} & = & 0\nonumber\\ 
\left.\partial_\sigma\left(\tilde{X}^2-B^{(\sigma)}_{21}\tilde{X}^1\right)\right|_{\sigma=0,\pi} & = & 0\,. 
\end{eqnarray}
The second equation defines a two dimensional hyper surface normal to a Dirichlet direction, which could be considered as the world-surface of a D-string. By linear combinations of two Neumann directions determined in the first, and the third equations, one can compose one temporal and one spatial coordinates on the world surface.
In all, a D-string will be slanted with an angle `$\pi/2-\arctan{B^{(\sigma)}_{12}}$' on the $(\tilde{X}^1$-$\tilde{X}^2)$-plane and move at a velocity `$-B^{(\sigma)}_{01}$' along $\tilde{X}^1$-direction. Since the fields $B^{(\sigma)}_{\mu\nu}$ are different at two boundaries $\sigma=0,\,\pi$, two D-strings have different tilting angles and different speed in general. Fig. 3 represents this. Interestingly enough, the intersection of two D-stings moves with just the velocity obtained  in Eq. (\ref{intervelo}). Consequently, we note that the strings pair-created over spacelike scissors keep pace with the intersection.

\subsubsection{a proposal to resolve the paradox}\label{6.5.2}

We are now faced with a very confusing situation. Two counter-intuitive results are involved with this. One is that how can the `would-be' on-shell strings move at the super-luminal speed? The other comes from the fact that the relation (\ref{in}) is invariant under the orientation flip that exchanges $(\vec{e},\,\vec{b})$ with $(\vec{e'},\,\vec{b'})$ (see Ref. \cite{cho03} for details). It implies that both the string and the anti-string will move at the same velocity. How can the string pair be separated to be on-shell? Below we propose a resolution for this dilemma.  
 
The case of anti-parallel electric fields but without magnetic fields could provide a good hint for the resolution. T-dual transformation along the direction of the electric fields results in a pair of parallel D- and (anti-)D-string, each of which recedes from each other at the speed $e$ and $e'$ respectively. (This is T-dual to D$0$-(anti-)D$0$-scattering discussed in Refs. \cite{bachas95,bachas98}.) The same configuration can be obtained from Fig. 3 by considering the limit of vanishing magnetic fields. However here, the ratio $a^2_0/a^0_0$ itself, though real valued, becomes very subtle because it is $+\infty$ or $-\infty$ depending on the sign of infinitesimal magnetic fields. This unreasonable result originates from our rough assumption of rigid boundary, $X^3(\sigma=0)=0$ and $X^3(\sigma=\pi)=y$, made in Eq. (\ref{bcboson}) and its T-dual. Although this `rigid rod ansatz' for the D-strings is quite good approximation in the weak string-coupling limit, it causes the notorious scissors paradox \cite{Taylor:1992}, because the inter-string keeping up with the intersection can be used as a super-luminal messenger. It is rather right to think that D-strings, having finite tension, will be bent by the inter-strings or else break the inter-strings. This possibility was discussed for the scissors configuration in Ref. \cite{bachas02}.

We pursue the idea further by considering the triple junction \cite{rey97} formed by the inter-strings and D-strings, expecting the inter-strings pull D-strings a bit, which prevents themselves from fleeing at the super-luminal speed.
\FIGURE{
\epsfbox{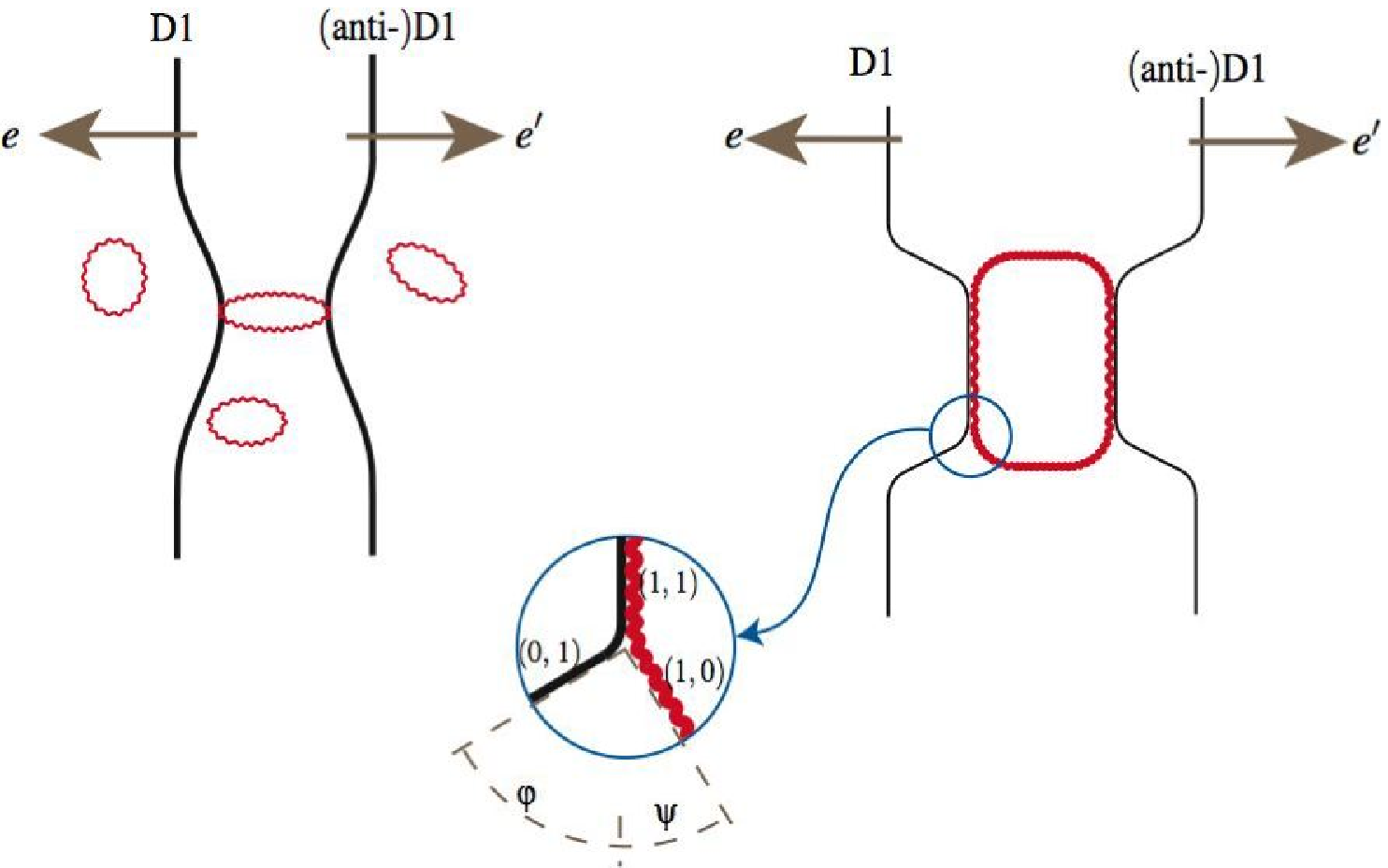}\label{fig4}
\caption{The pair created inter-strings retard D-strings to stop finally. In type IIB theory, the fundamental strings and D-strings form $SU(1, 1)$ doublet and are classified according to their NS-NS and R-R charges as $(p,\,q)$-strings. At each triple junction, their different tensions are balanced by the adjustment of the joining angles of three $(p,\,q)$-strings. The left cartoon describes the situation at weak string coupling. The inter-strings are detached from the moving D-strings to form closed strings. As the string coupling gets stronger (right figure), the inter-strings will bend D-strings more.}
}
Indeed the inter-strings will make triple junction with D-strings, as shown in Fig. 4. The angles $\psi$ and $\varphi$ are complementary to each other and are determined by the balancing condition for the tensions $T_{(p,q)}=T_{\text{eff}}\sqrt{p^2+(q/g_s)^2}$ (see \cite{rey97} for details). 
\begin{equation}
\cos{\varphi}=\frac{T_{(0,1)}}{T_{(1,1)}},\qquad \cos{\psi}=\frac{T_{(1,0)}}{T_{(1,1)}}
\end{equation}
As the string coupling becomes strong ($g_s\gg 1$), the angle $\varphi$ approaches the right angle, while in the weak coupling limit ($g_s\ll 1$), the angle $\psi$ goes to $\pi/2$. D-strings will be bent more by the strings created in pair. In general, pair creation process slows down D-strings because the kinetic energy of D-strings will be transferred to the inter-strings increasing their length. Obviously the increment in the length is not sufficient for the inter-strings to overcome the tension balance at the junction so that they slide over D-strings freely.  

As a side remark, we mention a consequence of the triple junction. This explains the `bending' of D-strings, which was proposed to resolve the paradox in Ref. \cite{bachas02}, in more quantitive way. There is an interesting mechanism of creating closed strings out of inter-strings. Due to the $(p, q)$ charge conservation, the only way of joining a pair of inter-strings with D-strings is to connect the inter-strings via $(1, 1)$ strings. In all, the picture look like a closed strings sandwiched by D-strings. 
In the weak coupling limit, it is more likely that the inter-strings go off D-strings to be closed strings. It corresponds to the `breaking' the inter-string described in \cite{bachas02}. This could be a possible decay mechanism of the unstable D-branes. 

The cases with arbitrary magnetic fields (Fig. 3) can be brought to the above case, as is shown in Ref. \cite{bachas02}. When the intersection is spacelike, D-strings can be made parallel by boosting. Indeed, in a frame boosted along the trajectory of the intersection with the speed 
\begin{equation}
\tanh\gamma=\frac{b+b'}{\sqrt{(e+e')^2+(e'b-eb')^2}}<1,
\end{equation}
both D-strings will be aligned vertically and recedes from each other. Especially when $b=b'$ and $e=e'$, both D-strings will move to the opposite direction along $\tilde{x}$-direction with the speed $\sqrt{e^2-b^2}<1$, in the frame boosted vertically with $\tanh\gamma=b/e < 1$.

Conclusively to say, though the junction makes spacelike motion, the center of mass of the pair-created inter-strings follows it at the speed $\tanh \gamma$, bending D-strings locally.
Especially in the weak coupling regime, the inter-string pair is more likely to be detached from D-strings to form a closed string. 

\FIGURE{
\epsfbox{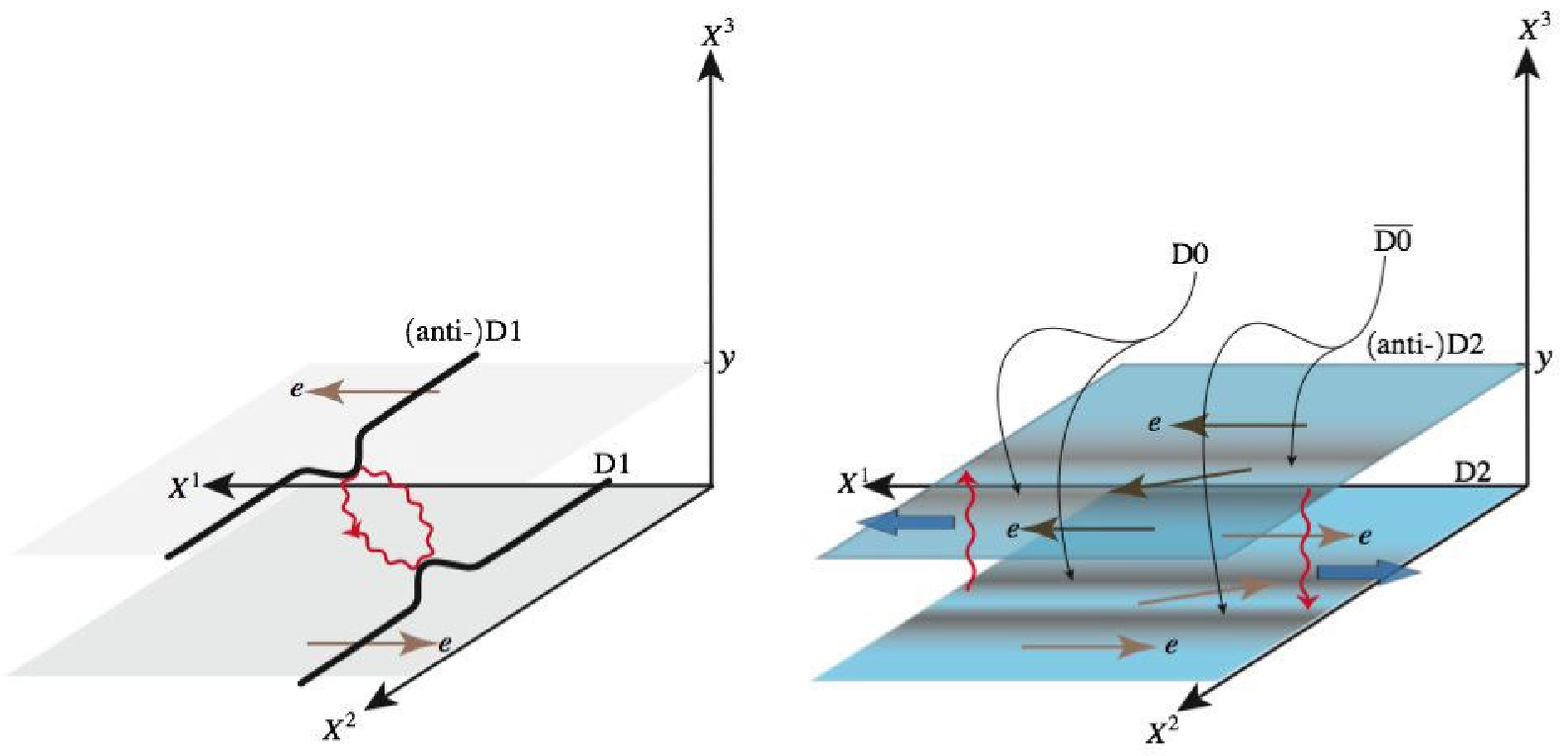}\label{fig5}
\caption{As (anti-)D-strings pass each other, the inter-strings are created in pair. Due to the tension balance at the triple junction, (anti-)D-strings are locally bent by the inter-strings. Under T-duality, this part corresponds to the local accumulation of D$0$-anti-D$0$ pairs around the current of inter-sting end points. This is just D-brane realization of Amp\`{e}re's law.}
}

\subsubsection{a conjecture about decaying or oscillatory electric fields}
Let us finally make some speculations about IIA physics. In IIA language, the D-string bending near the inter-string pair corresponds to the local accumulation of D0-$\overline{\text{D}0}$ pairs (Fig. 5) because the local tilting of D-strings caused by the `bending' is T-dual to the magnetic flux localized along the trails of the inter-strings. This could be the brane realization of Amp\`{e}re's law; magnetic field is induced around the current. 

The slow-down of D-string motion is T-dual to the decay of the electric fields. This looks natural because the inter-strings created in pair out of vacuum will move in the opposite direction under the influence of the electric fields. As the inter-strings, i.e., the charge carriers, accumulate on the world-volume according to the sign of their charges, the electric fields will be screened to diminish. 

As for D$1$-D$1$ cases, as D-strings slow down their motion, the whole configuration approaches a BPS state. This is dual to D$2$-D$2$ branes over which the anti-parallel electric fields decay to vanish, thus become stabilized. Meanwhile, in D$1$-anti-D$1$ system, even when (anti-)D-strings stop their motion, there is static attractive force between them due to their opposite Ramond-Ramond charges, which will make them move back and repeat the process. All these are T-dual to D$2$-anti-D$2$ system over which the electric fields make damping oscillation. We postpone the details to the future works.

\bigskip

\acknowledgments
We thank Costas Bachas, Kimyeong Lee, and Hyeonjoon Shin for valuable comments and discussions.
JHC is supported in part by KRF through Project No. KRF-2003-070-C00011. PO's work is partially supported by KRF grant KRF-2004-042-10682-0.

\begin{appendix}
\section{Quantization}\label{appi}
In order to invert the component matrix $(\Omega)_{IJ}$ of the symplectic form, we need to rewrite it in terms of the independent variables. The constraints, (\ref{in}) and (\ref{in2}), leave only three kinds of variables;
\begin{eqnarray}
(a^\mu_n) & = & (-\frac{b+b'}{e\sin\theta+e'\sin\theta'},\,0,\,1)\,a^2_n\equiv v\,a^2_n,\qquad (\mu=0,\,1,\,2)\nonumber \\
(a^\mu_{n+i\epsilon}) & = & (1,\,\frac{r+is}{p+iq},\,\frac{t+iu}{p+iq})\,a^0_{n+i\epsilon}\equiv w\,a^0_{n+i\epsilon},\nonumber\\
(\bar{a}^\mu_{-n-i\epsilon})&=& (1,\,\frac{r-is}{p-iq},\,\frac{t-iu}{p-iq})\,\bar{a}^0_{-n-i\epsilon}\equiv \bar{w}\,\bar{a}^0_{-n-i\epsilon},
\end{eqnarray}
and similar results for the fermionic variables. As a consequence, the symplectic forms in Eq. (\ref{sym}) reduce to
\begin{eqnarray}
\Omega_{B} &=&
-\frac{1}{4\pi\alpha^{\prime}}\Big(B^{(0)}-B^{(\pi)}\Big)_{\mu\nu}
\delta x^{\mu}\wedge\delta x^{\nu} - \frac{1}{\alpha^{\prime}}
\Big(\eta-B^{(\pi)}\eta^{-1}B^{(0)}\Big)_{\mu\nu}v^\nu\delta x^{\mu}\wedge\delta 
a_{0}^{2} \nonumber\\
&&+ \displaystyle\sum_{n\ne0}\frac{i}{n\alpha^{\prime}}\Big(\eta-B^{(0)}
\eta^{-1}B^{(0)}\Big)_{\mu\nu}v^\mu v^\nu \delta a_{-n}^{2}\wedge\delta a_{n}^{2} \nonumber\\
&&+ \displaystyle\sum_{n}\frac{i}{2(n+i\epsilon)\alpha^{\prime}}\Big(\eta-B^{(0)}
\eta^{-1}B^{(0)}\Big)_{\mu\nu}\bar{w}^\mu w^\nu \delta \bar{a}_{-n-i\epsilon}^{0}\wedge
\delta a_{n+i\epsilon}^{0}\,, \nonumber\\
\Omega_{F} &=& \sum_{r}\frac{i}{2\alpha^{\prime}}\Big(\eta-B^{(0)}\eta^{-1}
B^{(0)}\Big)_{\mu\nu} \bigg[v^\mu v^\nu\delta h_{-r}^{2}\wedge\delta h_{r}^{2} 
+ \frac{\bar{w}^\mu w^\nu}{2}\delta \bar{h}_{-r-i\epsilon}^{0}\wedge\delta h_{r+i\epsilon}^{0} 
\bigg] \,.
\label{sym2}
\end{eqnarray}
Now, taking the inverse of the component matrix of this reduced symplectic form, one obtains the Poisson algebra. Since we have worked out all the constraints, the Poisson brackets in the result are actually Dirac brackets. Standard quantization rule of replacing Dirac brackets with $i$ times (anti-)commutators lead us to the following algebra.
\begin{eqnarray}\label{commutator}
&&\left[x^{0}~,~x^{1}\right] = -\frac{2i\pi\alpha^{\prime}\Big(e^{\prime}
(1-bb^{\prime}-ee^{\prime}\cos\theta\cos\theta')\sin\theta' 
+ e(1+b^{\prime 2}-ee^{\prime}\cos\theta\cos\theta')\sin\theta\Big)}
{\Big(1-bb^{\prime}-ee^{\prime}\cos(\theta+\theta')\Big)^{2}\beta^{2}} 
\,, \nonumber\\
&&\left[x^{0}~,~x^{2}\right] = -\frac{2i\pi\alpha^{\prime}e\cos\theta
\Big(-b^{\prime}(b+b^{\prime})+e^{\prime 2}\sin^{2}\theta'
+ee^{\prime}\sin\theta\sin\theta'\Big)}
{\Big(1-bb^{\prime}-ee^{\prime}\cos(\theta+\theta')\Big)^{2}\beta^{2}}
\,, \nonumber\\
&&\left[x^{1}~,~x^{2}\right] = \frac{2i\pi\alpha^{\prime}\Big(-(b+b^{\prime})
+ee^{\prime}(b+b^{\prime})\cos\theta\cos\theta'+be^{\prime 2}\sin^{2}
\theta'-ee^{\prime}b^{\prime}\sin\theta\sin\theta'\Big)}
{\Big(1-bb^{\prime}-ee^{\prime}\cos(\theta+\theta')\Big)^{2}\beta^{2}}
\,, \nonumber\\
&&\left[a_{0}^{2}~,~x^{0}\right] = -\frac{i\alpha^{\prime}(b+b^{\prime})
(e\sin\theta+e^{\prime}\sin\theta')}
{\Big(1-bb^{\prime}-ee^{\prime}\cos(\theta+\theta')\Big)^{2}\beta^{2}}
\,, \nonumber\\
&&\left[a_{0}^{2}~,~x^{1}\right] = 0 \,, \nonumber\\
&&\left[a_{0}^{2}~,~x^{2}\right] = \frac{i\alpha^{\prime}(e\sin\theta
+e^{\prime}\sin\theta')^{2}}
{\Big(1-bb^{\prime}-ee^{\prime}\cos(\theta+\theta')\Big)^{2}\beta^{2}}
\,, \nonumber\\
&&\left[a_{m}^{2}~,~a_{n}^{2}\right] = -\frac{\alpha^{\prime}(e\sin\theta
+e^{\prime}\sin\theta')^{2}}
{2\Big(1-bb^{\prime}-ee^{\prime}\cos(\theta+\theta')\Big)^{2}\beta^{2}}
~ m\delta_{m+n} \,, \nonumber\\
&&\left[\bar{a}_{m-i\epsilon}^{0}~,~a_{n+i\epsilon}^{0}\right] = 
-\frac{2\alpha^{\prime}(p^{2}+q^{2})}{\Lambda} ~ (m-i\epsilon)\delta_{m+n} 
\,, \nonumber\\
&&\{h_{r}^{2}~,~h_{s}^{2}\} = -\frac{\alpha^{\prime}(e\sin\theta
+e^{\prime}\sin\theta')^{2}}
{\Big(1-bb^{\prime}-ee^{\prime}\cos(\theta+\theta')\Big)^{2}\beta^{2}}
~ \delta_{r+s} \,, \nonumber\\
&&\{\bar{h}_{r-i\epsilon}^{0}~,~h_{s+i\epsilon}^{0}\} = 
-\frac{4\alpha^{\prime}(p^{2}+q^{2})}{\Lambda} ~ \delta_{r+s} \,,
\end{eqnarray}
where $\Lambda = (p^{2}+q^{2})(1-e^2)-(1+b^{2})(r^{2}+s^{2}+t^{2}
+u^{2})-2be((pr+qs)\cos\theta+(pt+qu)\sin\theta)+e^{2}((t\cos\theta-r\sin\theta)^2+(u\cos\theta-s\sin\theta)^2)$ and $p^{2}+q^{2}$ are negative for the cases we are considering in this paper.

In order to make it in the canonical form, we redefine the variables 
taking the following nomalizations;
\begin{eqnarray}\label{normalization}
\alpha_{ n} &=& \sqrt{-\frac{2\left(1-bb^{\prime}-ee^{\prime}\cos(\theta
+\theta')\right)^{2}\beta^{2}} {\alpha^{\prime}(e\sin\theta
+e^{\prime}\sin\theta')^{2}}} ~ a_{ n}^{2} \,, \nonumber\\
\alpha_{ n+ i\epsilon} (\bar{\alpha}_{ n- i\epsilon})
&=& \sqrt{\frac{\Lambda}{2\alpha^{\prime}(p^{2}+q^{2})}} ~ 
a_{ n+ i\epsilon}^{0} (\bar{a}_{ n- i\epsilon}^{0}) \,, \nonumber\\
\varphi_{ r} &=& \sqrt{-\frac{\left(1-bb^{\prime}-ee^{\prime}\cos(\theta
+\theta')\right)^{2}\beta^{2}} {\alpha^{\prime}(e\sin\theta
+e^{\prime}\sin\theta')^{2}}} ~ h_{ r}^{2} \,, \nonumber\\
\varphi_{ r+ i\epsilon} (\bar{\varphi}_{ r- i\epsilon})
&=& \sqrt{\frac{\Lambda}{4\alpha^{\prime}(p^{2}+q^{2})}} ~ 
h_{ r+ i\epsilon}^{0} (\bar{h}_{ r- i\epsilon}^{0}) \,,
\end{eqnarray}
the commutators can be written in the canonical fashion
\begin{eqnarray}
\left[\alpha_{m}~,~\alpha_{n}\right] = m\delta_{m+n} \,, ~~~~~~~~~~ 
\left[\bar{\alpha}_{m-i\epsilon}~,~\alpha_{n+i\epsilon}\right] 
= -(m-i\epsilon)\delta_{m+n} \,, \nonumber\\
\{\varphi_{r}~,~\varphi_{s}\} = \delta_{r+s} \,, ~~~~~~~~~~ 
\{\bar{\varphi}_{r-i\epsilon}~,~\varphi_{s+i\epsilon}\}
= -\delta_{r+s} \,,
\end{eqnarray}
where the integer modes and the non-integer modes (involving $i\epsilon$) pertain to the direction $X^2$ of the transverse dimension and the longitudinal direction, respectively.

\section{Hamiltonian}\label{appii}

Let us look into the Hamiltonian for the brane world-volume directions.
\begin{eqnarray}
H_{(0,1,2)} & = & \frac{1}{4\pi\alpha'}\sum_{\mu=0,1,2}\int^\pi_0 d\sigma \left(\partial_\tau X^\mu\partial_\tau X_\mu +\partial_\sigma X^\mu\partial_\sigma X_\mu +i\psi^\mu_+\partial_\sigma\psi_{\mu+}-i\psi^\mu_-\partial_\sigma\psi_{\mu-}\right) \nonumber\\
 & = & \frac{1}{\alpha'}\sum_{\mu,\nu=0,1,2}\left(\eta-B^{(0)}\eta^{-1}B^{(0)}\right)_{\mu\nu}\left[\sum_{n}a^\mu_n\,a^\nu_{-n}+\frac{1}{4}\sum_{n}\left(a^\mu_{n+i\epsilon}\bar{a}^\nu_{-n-i\epsilon}+\bar{a}^\mu_{-n-i\epsilon}a^\nu_{n+i\epsilon}\right)\right.\nonumber\\
 &&\qquad\qquad\left.-\sum_{r}\frac{r}{2} h^\mu_r\,h^\nu_{-r}-\sum_{r}\frac{r+i\epsilon}{8}\left(h^\mu_{r+i\epsilon}\bar{h}^\nu_{-r-i\epsilon}-\bar{h}^\mu_{-r-i\epsilon}h^\nu_{r+i\epsilon}\right)\right].
\end{eqnarray}
In the second equality, we used the expressions of mode expansion, (\ref{solbo}) and (\ref{solfer}).  
Due to the constraints, (\ref{in}) and (\ref{in2}), the components of the modes are mutually dependent. In terms of the independent modes, the Hamiltonian reduces to 
\begin{eqnarray}
H_{(0,1,2)}&=&\frac{1}{2}\sum_{n}\alpha_n\,\alpha_{-n}-\frac{1}{2}\sum_{n}\left(\alpha_{n+i\epsilon}\bar{\alpha}_{-n-i\epsilon}+\bar{\alpha}_{-n-i\epsilon}\alpha_{n+i\epsilon}\right)\nonumber\\
&&-\sum_{r}\frac{r}{2}\varphi_r\,\varphi_{-r}+\sum_{r}\frac{r+i\epsilon}{2}\left(\varphi_{r+i\epsilon}\bar{\varphi}_{-r-i\epsilon}-\bar{\varphi}_{-r-i\epsilon}\varphi_{r+i\epsilon}\right)\nonumber\\
&=&\frac{1}{2}\alpha_0\,\alpha_0+\sum_{n>0}\alpha_{-n}\,\alpha_{n}-\sum_{n\ge 0}\bar{\alpha}_{-n-i\epsilon}\,\alpha_{n+i\epsilon}-\sum_{n> 0}\alpha_{-n+i\epsilon}\,\bar{\alpha}_{n-i\epsilon}\nonumber\\
&&+\sum_{r>0}r\varphi_{-r}\,\varphi_{r}-\sum_{r\ge 0}(r+i\epsilon)\bar{\varphi}_{-r-i\epsilon}\,\varphi_{r+i\epsilon}-\sum_{r> 0}(r-i\epsilon)\varphi_{-r+i\epsilon}\,\bar{\varphi}_{r-i\epsilon}\\
&&+\frac{1}{2}\left(\sum_{n>0}n+\sum_{n\ge 0}(n+i\epsilon)+\sum_{n> 0}(n-i\epsilon)\right)\nonumber\\
&&-\frac{1}{2}\left(\sum_{r>0}r+\sum_{r\ge 0}(r+i\epsilon)+\sum_{r>0}(r-i\epsilon)\right)\nonumber\\
&\equiv&\frac{1}{2}\alpha_0\,\alpha_0+N^\alpha+\bar{N}^\alpha_{\epsilon}+N^\alpha_{\epsilon}+N^\varphi+\bar{N}^\varphi_\epsilon+N^\varphi_\epsilon+({\cal E}^\alpha+{\cal E}^{\|}_X)+({\cal E}^\varphi+{\cal E}^{\|}_\psi).\nonumber
\end{eqnarray} 
The second equality rewrites the Hamiltonian in the normal ordered form. The final equality defines various number operators and zero-point energies. Making use of the well known `$\zeta$-function regularization'
\begin{equation}
\label{zeta}
\sum^{\infty}_{n=0}(n+\theta)=\frac{1}{24}-\frac{1}{8}(2\theta+1)^2,
\end{equation}
one can especially obtain
\begin{eqnarray}
\label{zpe}
{\cal E}^{\|}_X&=&\frac{1}{2}\sum_{n\ge 0}(n+i\epsilon)+\frac{1}{2}\sum_{n> 0}(n-i\epsilon)\nonumber\\
&=&\frac{i\epsilon}{2}(1-i\epsilon)-\frac{1}{12}\\&&\nonumber\\
{\cal E}^{\|}_\psi&=&
-\frac{1}{2}\sum_{r\ge 0}(r+i\epsilon)-\frac{1}{2}\sum_{r>0}(r-i\epsilon)\nonumber\\
& = & \left\{\begin{array}{ll}
    \displaystyle  \frac{1}{12}-\frac{i\epsilon}{2}(1-i\epsilon)&  \quad\text{(R)}  \\
    {}&{}\\
    \displaystyle -\frac{1}{24}+\frac{(i\epsilon)^2}{2}  &  \quad\text{(NS)} 
\end{array}\right. 
\end{eqnarray}
The part involving integer modes are in the conventional form and composes the transverse part (denoted by $\bot$) together with the other dimensions normal to the branes. 

\section{Momentum Constraints under T-duality}\label{appiii}

In this section, we show that the constraints, (\ref{in}) and (\ref{in2}), among the `momentum' components are invariant under T-duality. T-duality along an isometric direction switches Dirichlet boundary condition to Neumann boundary condition and vice versa. More specifically, T-duality along the direction $X^1$ deforms the terms in the boundary condition at $\sigma=0,\,\pi$ as
\begin{eqnarray}
\label{tdual}
&&\partial_{+}X^1 \leftrightarrow \partial_{+}\tilde{X}^1\nonumber\\
&&\partial_{-}X^1 \leftrightarrow -\partial_{-}\tilde{X}^1.
\end{eqnarray}
This affects the solution as
\begin{eqnarray}
X^1(\sigma^+,\,\sigma^-)=X^1(\sigma^+)+X^1(\sigma^-) &\leftrightarrow& \tilde{X}^1(\sigma^+,\,\sigma^-)=X^1(\sigma^+)-X^1(\sigma^-). 
\end{eqnarray}
Under this T-duality, the solution (\ref{solbo}) satisfying the boundary condition at $\sigma=0$ transforms to
\begin{equation}
\label{soliib}
\tilde{X}^\mu(\sigma^+,\,\sigma^-)=X^\mu(\sigma^+)+T^\mu{}_\nu X^\nu(\sigma^-),
\end{equation}
where we introduced `T-dualizing matrix', $(T^\mu{}_\nu)=\text{diag}(1,\,-1,\,1)$. This transformed solution satisfies the boundary condition 
\begin{equation}
\label{bciib0}
E^{(\sigma)}_{\nu\mu}\partial_+\tilde{X}^\nu=E^{(\sigma)}_{\mu\nu}T^\nu{}_\rho\partial_-\tilde{X}^\rho
\end{equation}
at $\sigma=0$. Since
\begin{eqnarray}
\partial_+\tilde{X}_\mu & = & E^{(0)}_{\mu\nu}\left[a^\nu_0+\sum_{n\ne 0}a^\nu_n e^{-in\sigma^+}+\sum^{\infty}_{n=\infty}\frac{1}{2}\left(a^\nu_{n+i\epsilon}e^{-i(n+i\epsilon)\sigma^+}+\bar{a}^\nu_{-n-i\epsilon}e^{i(n+i\epsilon)\sigma^+}\right)\right]\,,\\
\partial_-\tilde{X}_\mu & = & T_{\mu}{}^{\rho}E^{(0)}_{\nu\rho}\left[a^\nu_0+\sum_{n\ne 0}a^\nu_n e^{-in\sigma^-}+\sum^{\infty}_{n=\infty}\frac{1}{2}\left(a^\nu_{n+i\epsilon}e^{-i(n+i\epsilon)\sigma^-}+\bar{a}^\nu_{-n-i\epsilon}e^{i(n+i\epsilon)\sigma^-}\right)\right]\,,\nonumber
\end{eqnarray}
the boundary condition is satisfied at $\sigma=\pi$, if
\begin{eqnarray}
&&E^{(\pi)}_{\rho\mu}E^{(0)\rho}{}_\nu\left[a^\nu_0+\sum_{n\ne 0}a^\nu_n e^{-in\sigma^+}+\sum^{\infty}_{n=\infty}\frac{1}{2}\left(a^\nu_{n+i\epsilon}e^{-i(n+i\epsilon)\sigma^+}+\bar{a}^\nu_{-n-i\epsilon}e^{i(n+i\epsilon)\sigma^+}\right)\right]_{\sigma=\pi}\nonumber\\ &&\quad=E^{(\pi)}_{\mu\kappa}T^{\kappa}{}_{\lambda}T^{\lambda\rho}E^{(0)}_{\nu\rho}\left[a^\nu_0+\sum_{n\ne 0}a^\nu_n e^{-in\sigma^-}+\sum^{\infty}_{n=\infty}\frac{1}{2}\left(a^\nu_{n+i\epsilon}e^{-i(n+i\epsilon)\sigma^-}+\bar{a}^\nu_{-n-i\epsilon}e^{i(n+i\epsilon)\sigma^-}\right)\right]_{\sigma=\pi}.\nonumber
\end{eqnarray}
As $T^{\kappa}{}_{\lambda}T^{\lambda\rho}=\eta^{\kappa\rho}$, the condition becomes exactly the same as the one used prior to T-duality.
\end{appendix}

\end{document}